\documentclass[aps,prx,preprint,superscriptaddress]{revtex4-2}
\usepackage{amsmath}
\usepackage{amssymb}

\newcommand{\betagao}{$\beta$-Ga$_2$O$_3$}

\usepackage{graphicx}
\usepackage{array}
\usepackage{siunitx}
\usepackage{dcolumn}
\usepackage[version=4]{mhchem}
\usepackage{svg}
\usepackage{gensymb}
\usepackage{booktabs}
\usepackage{ragged2e}
\usepackage{threeparttable} 
\usepackage{tabularx}
\usepackage{textcomp}
\usepackage{color}
\usepackage{xurl}
\usepackage{hyperref}

\begin{document}

\title{Nanoscale origin of thermal anisotropy in monoclinic \texorpdfstring{$\beta$-Ga$_2$O$_3$}{beta-Ga2O3}}

\author{Kai Xu}
\affiliation{Institut de Ci\`{e}ncia de Materials de Barcelona, (ICMAB-CSIC), 08193 Cerdanyola del Vallès, Spain}

\author{Shuo Zhao}
\affiliation{The NOMAD Laboratory at BIFOLD, Technische Universit\"at Berlin, Franklinstraße 28/29, 10587 Berlin, Germany}

\author{Luca Sung-Min Choi}
\author{Nils Bernhardt}
\affiliation{Technische Universit\"at Berlin, Institut f\"ur Physik und Astronomie, Hardenbergstraße 36, 10623 Berlin, Germany}

\author{Pouria Emtenani}
\affiliation{Paul-Drude-Institut f\"ur Festk\"orperelektronik, Leibniz-Institut im Forschungsverbund Berlin e.V., Hausvogteiplatz 5–7, 10117 Berlin, Germany}
\affiliation{Technische Universit\"at Berlin, Institut f\"ur Physik und Astronomie, Hardenbergstraße 36, 10623 Berlin, Germany}
\author{Felix Nippert}
\affiliation{Technische Universit\"at Berlin, Institut f\"ur Physik und Astronomie, Hardenbergstraße 36, 10623 Berlin, Germany}

\author{Moritz Meißner}
\affiliation{Paul-Drude-Institut f\"ur Festk\"orperelektronik, Leibniz-Institut im Forschungsverbund Berlin e.V., Hausvogteiplatz 5–7, 10117 Berlin, Germany}

\author{Hans Tornatzky}
\affiliation{Paul-Drude-Institut f\"ur Festk\"orperelektronik, Leibniz-Institut im Forschungsverbund Berlin e.V., Hausvogteiplatz 5–7, 10117 Berlin, Germany}

\author{Oliver Bierwagen}
\affiliation{Paul-Drude-Institut f\"ur Festk\"orperelektronik, Leibniz-Institut im Forschungsverbund Berlin e.V., Hausvogteiplatz 5–7, 10117 Berlin, Germany}

\author{Piero Mazzolini}
\affiliation{Department of Mathematical, Physical and Computer Sciences, University of Parma, Viale delle Scienze 7/A, 43124 Parma, Italy}

\author{Riccardo Mincigrucci}
\affiliation{Elettra Sincrotrone Trieste S.C.p.A., Strada Statale 14, km 163.5, 34149 Basovizza (TS), Italy}

\author{Laura Foglia}
\affiliation{Elettra Sincrotrone Trieste S.C.p.A., Strada Statale 14, km 163.5, 34149 Basovizza (TS), Italy}

\author{Danny Fainozzi}
\affiliation{Institute of Applied Physics, University of Bern, CH-3012 Bern, Switzerland}

\author{Filippo Bencivenga}
\affiliation{Elettra Sincrotrone Trieste S.C.p.A., Strada Statale 14, km 163.5, 34149 Basovizza (TS), Italy}

\author{Zbigniew Galazka}
\affiliation{Leibniz-Institut für Kristallzüchtung, Max-Born-Straße 2, 12489 Berlin, Germany}

\author{George Fytas}
\affiliation{Max Planck Institute for Polymer Research, Ackermannweg 10, 55128 Mainz, Germany}
\affiliation{Faculty of Physics, Adam Mickiewicz University, Uniwersytetu Poznanskiego 2, 61-614 Poznan, Poland}

\author{Bartlomiej Graczykowski}
\affiliation{Max Planck Institute for Polymer Research, Ackermannweg 10, 55128 Mainz, Germany}
\affiliation{Faculty of Physics, Adam Mickiewicz University, Uniwersytetu Poznanskiego 2, 61-614 Poznan, Poland}

\author{Riccardo Rurali}
\affiliation{Institut de Ci\`{e}ncia de Materials de Barcelona, (ICMAB-CSIC), 08193 Cerdanyola del Vallès, Spain}

\author{Matthias Scheffler}
\affiliation{The NOMAD Laboratory at BIFOLD, Technische Universit\"at Berlin, Franklinstraße 28/29, 10587 Berlin, Germany}

\author{Christian Carbogno}
\email{carbogno@fhi-berlin.mpg.de}
\affiliation{The NOMAD Laboratory at BIFOLD, Technische Universit\"at Berlin, Franklinstraße 28/29, 10587 Berlin, Germany}
\affiliation{Theory Department, Fritz Haber Institute of the Max Planck Society, Faradayweg 4-6, D-14195 Berlin, Germany}

\author{Juan Sebastián Reparaz}
\email{jsreparaz@icmab.es}
\affiliation{Institut de Ci\`{e}ncia de Materials de Barcelona, (ICMAB-CSIC), 08193 Cerdanyola del Vallès, Spain}

\author{Markus R. Wagner}
\email{wagner@pdi-berlin.de}
\affiliation{Paul-Drude-Institut f\"ur Festk\"orperelektronik, Leibniz-Institut im Forschungsverbund Berlin e.V., Hausvogteiplatz 5–7, 10117 Berlin, Germany}
\affiliation{Technische Universit\"at Berlin, Institut f\"ur Physik und Astronomie, Hardenbergstraße 36, 10623 Berlin, Germany}

\begin{abstract}
As wide-bandgap power electronics scale toward higher power densities and smaller dimensions, the length-scale dependence of phonon-mediated heat transport becomes increasingly important. Monoclinic $\beta$-Ga$_2$O$_3$ is a promising ultra-wide-bandgap semiconductor for power electronics, but its low and anisotropic thermal conductivity limits heat removal, making it a compelling model system for nanoscale thermal transport studies. Here, we demonstrate a pronounced nanoscale enhancement of thermal anisotropy and uncover the underlying phonon mechanism by combining complementary thermal and acoustic spectroscopy techniques with first-principles calculations. Anisotropic thermoreflectance thermometry determines the full thermal-conductivity tensor from angular measurements within the (100), (010), and (001) crystallographic planes. Angle-resolved Brillouin light scattering reveals that low-frequency acoustic phonons exhibit only weak velocity anisotropy, insufficient to account for the measured thermal anisotropy. Frequency-resolved \textit{ab initio} Green--Kubo calculations show that the bulk anisotropy pattern emerges only when the accumulated conductivity includes phonon modes above $\sim$4~THz, whereas the sub-3~THz acoustic contributions are nearly isotropic. Extreme-ultraviolet transient thermal grating spectroscopy reveals that reducing the grating period to $28$~nm strongly enhances the in-plane thermal anisotropy, most markedly in the (001) plane, where it reaches nearly twice its macroscopic value. Reconstructed and calculated mean-free-path (MFP) accumulation functions link this enhancement to the stronger directional contrast of short-MFP phonons. Together, these results identify high-frequency, short-MFP phonons as the dominant origin of thermal anisotropy in $\beta$-Ga$_2$O$_3$, while long-MFP, low-frequency phonons attenuate the directional contrast at macroscopic scales. These findings establish that thermal anisotropy depends on the characteristic transport length, providing a design parameter for nanoscale heat management in low-symmetry materials and ultra-wide-bandgap electronics.
\end{abstract}

\maketitle

The crystal symmetry of materials is decisive in determining their physical properties. Of the seven crystal lattice systems, the monoclinic structure has the second lowest symmetry, with the monoclinic plane defined by non-orthogonal crystal axes. A consequence of the low crystal symmetry is the directional dependence (anisotropy) of physical properties such as the dielectric constant, elasticity, and thermal conductivity~\cite{Newnham2004}. This low symmetry gives rise to unconventional phenomena such as hyperbolic shear polaritons~\cite{passler_hyperbolic_2022,carini_spectral_2026}, charge density waves or topological phase transitions not observed in higher-symmetry materials~\cite{basov_towards_2017}.

Among the monoclinic crystals, \betagao{} plays a unique role for future power electronic devices as an ultra-wide bandgap semiconductor with widely tunable electrical conductivity, high electrical breakdown field, and availability of large-area substrates for homoepitaxial growth~\cite{Pearton,Pearton2018,Higashiwaki,Higashiwaki2,Higashiwaki2017,Kokubuna2007,Oshima2008,Spencer2022,Green2022}. Its monoclinic structure leads to a profound anisotropy of its optical~\cite{Matsumoto1974,Ueda1997,Yamaguchi2004,Sturm2015,Varley2015,Onuma2015,Onuma2016,Onuma2016b,Furthmueller2016,schubert_anisotropy_2016,Kranert2016SciRep,Kranert2016PRL,Sturm2016b,Ricci2016,Grundmann2017,Mock2017,Ratnaparkhe2017,Gopalan2020,Cho2021,Zhang2021optical,Spencer2022,meissner:2024anisotropy_PLE_bGaO,hajizadeh_selftrapped_2026,tornatzky_complete_2026}, elastic~\cite{Furthmueller2016,adachi2018unusual,Grashchenko2018,Luan2019,grundmann2020principal}, and thermal~\cite{Guo2015,Handwerg2016,Slomski2017,Mu2017,Jiang2018,Zhang2020,Galazka2022,Klimm2022,Ahrling2024,guo_revisiting_2026} properties, creating both opportunities and challenges for device engineering. In particular, the highly anisotropic and low thermal conductivity of \betagao{} poses a major challenge for its technological breakthrough in power electronic applications. While several works have investigated the thermal conductivity of \betagao{}, the reported values for the anisotropy ratios between the different crystal directions vary by more than 50\%, and no coherent picture for the phononic origin of the anisotropy has yet emerged. 

Numerous experimental~\cite{villora_electrical_2008,Galazka2014,Guo2015,Handwerg2015,Handwerg2016,Slomski2017,Mu2017,Jiang2018,Zhang2020,Galazka2022,Klimm2022,Ahrling2024,guo_revisiting_2026} and theoretical~\cite{Santia2015,Yan2018,Li2020,Sun2023,Chen2023,Yang2024,wang_dissimilar_2024,Rybin2024,Safieddine2024,Tu2025,zhou_abnormal_2025,alkandari_anisotropic_2025,su_unveiling_2026} studies have investigated the thermal conductivity $\kappa$ of \betagao{}, with reported values ranging from 9.4 to 29.6 Wm$^{-1}$K$^{-1}$ across the lowest to highest $\kappa$ directions at room temperature. Most studies, both theoretical and experimental, consistently indicate an ordering of thermal conductivities across crystal directions, specifically $\kappa_{[010]} > \kappa_{[001]} > \kappa_{[100]}$. However, there remains considerable inconsistency in the absolute values reported across these studies, as well as in the interpretation of their nanoscale origin. A summary of representative experimental and theoretical values is provided in Table~S2 of the Supporting Information (SI).

Experimental methods employed to obtain $\kappa$ for each crystallographic direction include the $2\omega$~\cite{Handwerg2016} and $3\omega$ techniques~\cite{Handwerg2015,Handwerg2016,Slomski2017,Ahrling2024}, laser flash methods~\cite{villora_electrical_2008,Galazka2014,Mu2017,Galazka2022,Klimm2022}, and time-domain thermoreflectance (TDTR) in its conventional~\cite{Guo2015,Zhang2020,guo_revisiting_2026}, beam-offset~\cite{Zhang2020}, and elliptical-beam~\cite{Jiang2018} implementations. However, these studies generally do not offer a comprehensive, self-consistent mapping of the angular dependence of thermal conductivity across all planes. Furthermore, frequency- and length-scale-dependent thermal transport anisotropy remains poorly characterized, and the thermal conductivity values derived from TDTR can be strongly affected by both the modulation frequency and the composition of the metal/$\beta$-Ga$_2$O$_3$ interface~\cite{guo_revisiting_2026}.

Complementing these experimental efforts, theoretical work has also advanced the understanding of the thermal conductivity of \betagao{}. 
By performing molecular dynamics~\cite{Li2020,Sun2023,wang_dissimilar_2024,zhou_abnormal_2025,su_unveiling_2026} or lattice dynamics simulations~\cite{Santia2015,Yan2018,Chen2023,Yang2024,Rybin2024,Safieddine2024,Tu2025,alkandari_anisotropic_2025} and describing the interatomic interactions either directly via density functional theory (DFT)~\cite{Santia2015,Yan2018,Chen2023,Yang2024,Safieddine2024,Tu2025,alkandari_anisotropic_2025} or via machine-learned interatomic potentials trained on DFT data~\cite{Li2020,Sun2023,wang_dissimilar_2024,Rybin2024,zhou_abnormal_2025},
theoretical models have offered valuable insights into the anisotropy and directional dependence of thermal conductivity, often aligning with and enriching experimental observations. These methods enable the exploration of frequency-dependent thermal transport, which is usually not investigated experimentally due to the lack of suitable methods. Nevertheless, previous theoretical studies have not yet provided a self-consistent description of the full thermal-conductivity tensor together with its angular, frequency-dependent, and length-scale–dependent anisotropy. Persistent discrepancies between theoretical predictions and measurements indicate that both experimental and computational approaches still require refinement for accurate characterization of nanoscale thermal anisotropy. 

We address this gap through a comprehensive, multi-technique approach to characterize the frequency- and length-scale-dependent thermal anisotropy of materials. Employing a set of complementary, non-contact, all-optical spectroscopic techniques, we systematically explore the thermal anisotropy of \betagao{}. The techniques utilized include anisotropic thermoreflectance thermometry (ATT)~\cite{Perez2022}, angle-resolved Brillouin light scattering (BLS)~\cite{kargar_advances_2021, wright_acoustic_2025}, and extreme UV transient thermal grating spectroscopy (EUV-TTG)~\cite{bencivenga_nanoscale_2019, bencivenga_extreme_2023}. The combination of these methods enables precise measurement of directional thermal transport across diverse frequency and length scales, thereby providing insights into the mechanisms driving thermal anisotropy. To gain deeper insights into the origin of the observed thermal anisotropy, we perform \textit{ab initio} Boltzmann Transport Equation (BTE) and fully anharmonic \textit{ab initio} Green-Kubo (aiGK) calculations, capturing intrinsic thermal properties and non-equilibrium phonon effects. 

\begin{figure*}[h!]
\includegraphics[width=\textwidth]{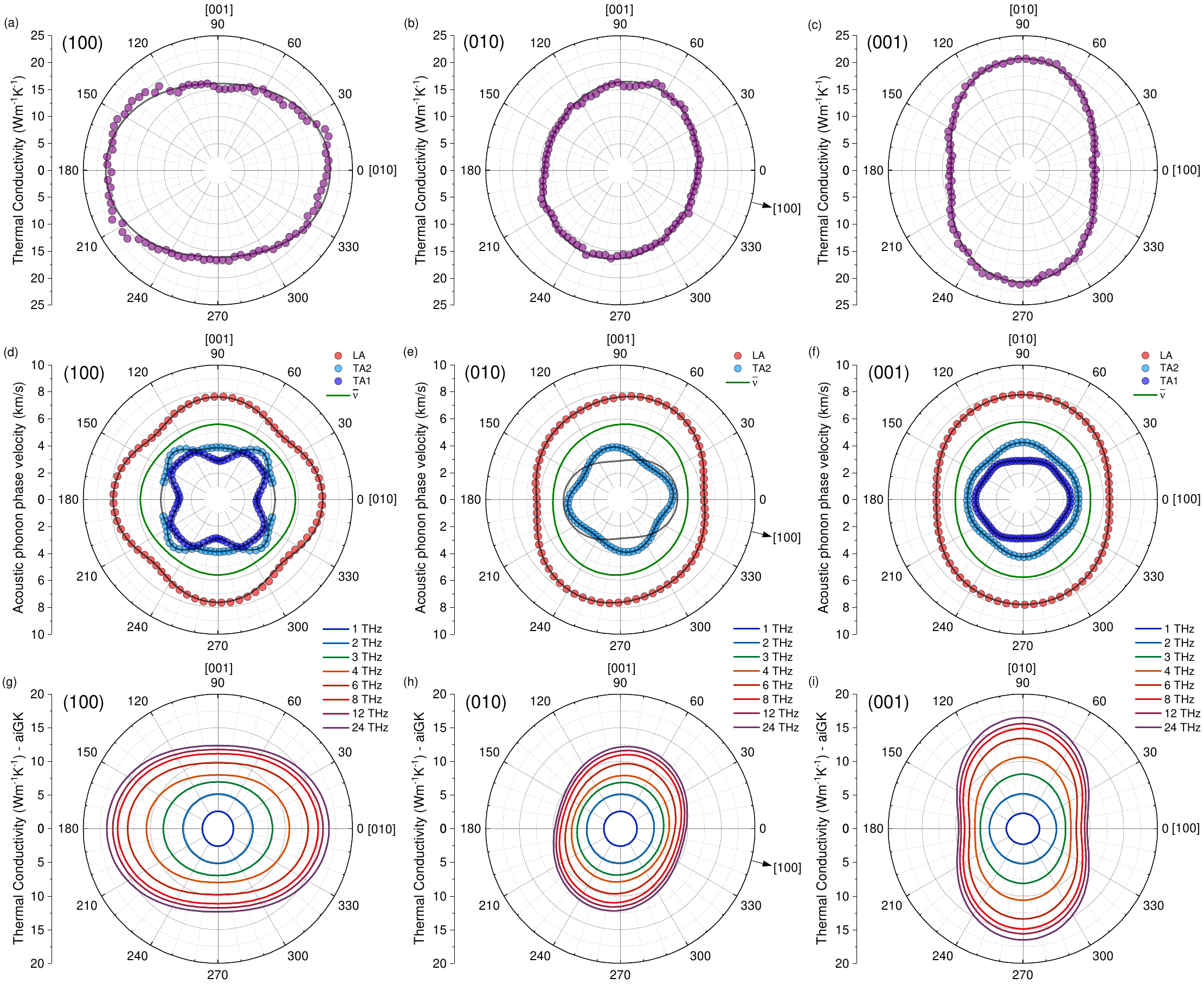}
\caption{(a)-(c)~Angular dependence of the thermal conductivity measured for the (100), (010), and (001) crystallographic planes of \betagao{}. The solid black lines are fits to the experimental data points (purple circles) measured at ambient conditions. (d)-(f) 
Angular dependence of the phase velocity of acoustic waves measured for the principal crystallographic planes. The full circles represent measured values  while solid lines are numerical fits of the experimental data. Symbols LA, TA2, and TA1 denote longitudinal, fast transverse and slow transverse bulk acoustic waves, respectively. (g)-(i) 
Angular dependence of the accumulated thermal conductivity of all three planes with frequency cut-offs between 1~THz and 24~THz as obtained by \textit{ab initio} Green-Kubo calculations.}
\label{fig:Fig1}
\end{figure*}

Figure~\ref{fig:Fig1} displays a direct comparison of the anisotropy of measured thermal and acoustic properties in conjunction with the calculated frequency dependence of phonon contributions to thermal conductivity across the three principal planes of \betagao{}. Using ATT (section S4 of the SI), we measured the angular dependence of thermal conductivity in the (100), (010), and (001) planes of \betagao{}, as shown in Figures~\ref{fig:Fig1}(a)-(c). By generating directional in-plane heat flow and performing quasi-continuous angular scans within each crystallographic plane, ATT is particularly sensitive to in-plane thermal anisotropy, allowing the anisotropy ratio to be directly extracted from fits to the polar plots. The thermal conductivities obtained along the principal directions were independently verified using beam-offset frequency-domain thermoreflectance with a one-dimensional optical heat source~\cite{xu_-plane_2023}. This recently developed method provides enhanced sensitivity to in-plane heat transport and yielded values consistent with the ATT measurements within experimental uncertainty (section S5 of the SI).

To fit the angular dependence measured by ATT, the equivalent scalar thermal conductivity along a direction $\mathbf{n}$ is expressed as the quadratic form $\kappa_{\mathrm{equiv}}(\mathbf{n}) = \mathbf{n}^{\mathrm{T}}\boldsymbol{\kappa}\mathbf{n}$, where $\boldsymbol{\kappa}$ is the thermal conductivity tensor. For a given crystallographic plane, treated as a two-dimensional Cartesian coordinate system $(x,y)$, the thermal conductivity along a direction forming an angle $\theta$ with the $x$-axis is $\kappa(\theta)=\kappa_{xx}\cos^{2}\theta+2\kappa_{xy}\sin\theta\cos\theta+\kappa_{yy}\sin^{2}\theta$.

The directional in-plane thermal conductivities measured in each crystallographic plane were fitted using this expression (black solid lines in Fig.~\ref{fig:Fig1}(a)-(c)), enabling the determination of the thermal conductivity tensor components and the corresponding in-plane anisotropies. The angular dependence is relatively flat over a broad range of directions around the principal crystallographic axes. This behavior is particularly evident near the [001] direction in the (100) plane and near the [100] direction in the (001) plane, where the comparatively high thermal conductivity along [010] produces elongated polar distributions. In the (010) plane, the angular distribution is slightly rotated relative to the principal crystallographic directions as the [100] and [001] directions are nonorthogonal in the monoclinic structure. The (001) plane exhibits the largest in-plane thermal anisotropy with $\kappa_{[010]}/\kappa_{[100]}=1.56$, followed by the (100) plane with $\kappa_{[010]}/\kappa_{[001]}=1.33$, whereas the (010) plane exhibits the smallest anisotropy with $\kappa_{[001]}/\kappa_{[100]}=1.18$. These values are in good qualitative agreement with previous reports~\cite{Guo2015,Jiang2018,Klimm2022,guo_revisiting_2026}. Additional measurements on Si-doped crystals show that the in-plane thermal-anisotropy ratios remain unchanged within experimental uncertainty over free-carrier concentrations from \(1\times10^{16}\) to \(5\times10^{18}\,\mathrm{cm^{-3}}\), indicating that impurity and free-carrier scattering do not measurably alter the anisotropic thermal transport in the investigated carrier-concentration range (section S6 of the SI).

To facilitate the description of the anisotropic thermal conductivity of \betagao{}, we introduce a symmetry-adapted orthonormal Cartesian coordinate system $(x_{11},x_{22},x_{33})$ such that $x_{22}\parallel[010]\parallel\textbf{\textit{b}}$ and $x_{33}\parallel[001]\parallel\textbf{\textit{c}}$, while the $[100]\parallel\textbf{\textit{a}}$ direction lies in the $x_{11}$--$x_{33}$ plane and forms an angle $\alpha=13.7^\circ$ with the $x_{11}$ axis, perpendicular to the (100) plane (section S1 of the SI). In this basis, the thermal conductivity tensor takes the form
\begin{equation}
\kappa_{ij} =
\begin{pmatrix}
\kappa_{11} & 0 & \kappa_{13}\\
0 & \kappa_{22} & 0\\
\kappa_{13} & 0 & \kappa_{33}
\end{pmatrix}.
\label{eq:kappa_tensor}
\end{equation}
The experimentally obtained components of the thermal conductivity tensor are listed in Table~\ref{thermal_table}, together with the theoretical values obtained from \textit{ab initio} Green-Kubo as well as BTE calculations (see Materials and Methods).
\begin{table}[t]
\caption{Comparison between experimental and theoretical values of the thermal conductivity tensor $\kappa_{ij}$ of a \betagao{} crystal.}
\begin{tabular}{cccc}
\toprule[1pt]
Tensor Element & Exp. (ATT)  & BTE  & aiGK\\
$\kappa_{ij}$         &  [Wm$^{-1}$K$^{-1}$] &  [Wm$^{-1}$K$^{-1}$] &  [Wm$^{-1}$K$^{-1}$]\\
\midrule[0.5pt]
$\kappa_{11}$ & $13.9\pm 1.0$ & 14.3 & $9.7\pm 0.2$\\
$\kappa_{22}$ & $21.2\pm 1.4$ & 22.6 & $16.5\pm 0.9$\\
$\kappa_{33}$ & $16.0\pm 1.1$ & 14.6 & $12.1\pm 0.2$\\
$\kappa_{13}$ & $0.8\pm 0.3$ & -- & $0.7\pm 0.1$\\
\bottomrule[1pt]
\end{tabular}
\label{thermal_table}
\end{table}

Notably, the theoretical predictions are generally in good agreement with the experimental results obtained in this work and the recent TDTR study by Guo \textit{et al.}~\cite{guo_revisiting_2026}, particularly for the diagonal tensor components and their relative ordering, This agreement persists even for the BTE calculations, which include only lowest-order anharmonicity, indicating that the dominant contribution to heat transport can be described by semi-classical phonon scattering. Taking into account additionally anisotropic lattice expansion and higher-order anharmonic effects in aiGK calculations~\cite{aiGK} reduces the absolute values of $\boldsymbol{\kappa}$ as expected from the inclusion of additional phonon scattering processes. The remaining quantitative differences between theory and experiment can arise from the sensitivity of $\boldsymbol{\kappa}$ to details of the interatomic interactions, including the employed description of exchange and correlation and the presence of impurities. The observed variation is consistent with high-level benchmark calculations~\cite{Knoop2023_PRL}. In contrast to the absolute conductivities, relative quantities such as anisotropy ratios are captured particularly well. For example, the aiGK value of $\kappa_{22}/\kappa_{33}=1.36$ is in excellent agreement with the experimental value of $1.33$ and the corresponding mean value of $1.36$ for the doping series in section S6 of the SI. These results support a consistent description of anisotropic thermal transport in \betagao{} across the atomistic and macroscopic levels.

To assess the contribution of direction-dependent acoustic phonon velocities to the observed thermal anisotropy, we examine the acoustic anisotropy in each crystallographic plane using angle-resolved Brillouin light scattering (BLS). Figures~\ref{fig:Fig1}(d)-(f) show the measured phase velocities of the longitudinal acoustic (LA) and two transverse acoustic (TA1 and TA2) bulk modes. The solid lines are calculated from the fitted elastic constants (section S2 of the SI). As shown in Figures~S2(a)-(c), the angular anisotropies measured by BLS closely reproduce those of the phase velocities calculated from \textit{ab initio} molecular dynamics simulations at 300\,K and used in the aiGK calculations. The angle-dependent cubic mean velocities
$\bar{v}(\phi) = \left[ \left( v_{L}^3(\phi) + v_{T1}^3(\phi) + v_{T2}^3(\phi) \right)/3 \right]^{\frac{1}{3}}$
\noindent are displayed in Figures~\ref{fig:Fig1}(d)-(f) as solid green lines (section S3 of the SI). We next examine the acoustic anisotropy of the three crystallographic planes in detail.

Figure~\ref{fig:Fig1}(d) shows that all acoustic modes propagating in the (100) plane are anisotropic. Although the overall angular dependence initially appears nearly fourfold symmetric, closer inspection reveals a two-fold symmetry, as the phase velocities of phonons propagating along [010] and [001] differ by 0.5--9\% for all three polarizations. This minor difference is also retained in the cubic mean velocities, with $\bar{v}_{[010]} = 5752$\,m/s and $\bar{v}_{[001]} = 5588$\,m/s, whereas $\bar{v}_{[100]} = 5020$\,m/s is substantially lower. For the (100) plane, the resulting acoustic anisotropy ratio between [010] and [001] is 1.03 and arises primarily from the transverse modes.

In the (010) plane [Fig.~\ref{fig:Fig1}(e)], the angular dispersion exhibits a distinct two-fold symmetry, with characteristic distortions caused by the monoclinic angle. The maxima and minima do not coincide with high-symmetry directions. The TA1 mode is absent from the BLS spectra in this plane because its photoelastic coupling vanishes for all light polarizations. The cubic mean velocities yield an anisotropy ratio of 1.11. For the (001) plane shown in Figure~\ref{fig:Fig1}(f), all three bulk acoustic modes are active in BLS. Their angular dispersion displays a clear two-fold symmetry and yields the largest in-plane anisotropy ratio of 1.15. As expected for a low-symmetry crystal, the acoustic phonon velocities exhibit pronounced anisotropy. However, their angular dependence differs substantially from that of the thermal conductivity shown in Figures~\ref{fig:Fig1}(a)-(c). To understand this difference, we examine the frequency-resolved contributions to $\boldsymbol{\kappa}$, which allows us to establish a connection between the phase velocities of acoustic phonons and the thermal conductivity. 

Since most contributions to $\kappa$ stem from particle-like, kinetic transport (see the discussion of Table~\ref{thermal_table}), we apply the Peierls--Boltzmann expression derived from kinetic theory,
\(\kappa^{\alpha\beta} = \sum_{\vec{q},j}
C_v(\vec{q},j)\,v_g^{\alpha}(\vec{q},j)\,
\Lambda^{\beta}(\vec{q},j)\),
where \(C_v(\vec{q},j)\) is the volumetric isochoric heat capacity of the phonon mode with wave vector \(\vec{q}\) and branch \(j\), \(v_g^{\alpha}(\vec{q},j)\) is the \(\alpha\)-component of its group velocity, and \(\Lambda^{\beta}(\vec{q},j) = v_g^{\beta}(\vec{q},j)\tau(\vec{q},j)\) is the corresponding \(\beta\)-component of the phonon mean free path (MFP), with \(\tau(\vec{q},j)\) denoting the phonon relaxation time. The sum extends over all phonon wave vectors \(\vec{q}\) and branches \(j\), reflecting the cumulative contributions of individual phonon modes to the macroscopic thermal conductivity tensor \(\kappa^{\alpha \beta}\). This expression shows that thermal conductivity is governed by the directional phonon group velocities and mean free paths, thereby linking mode-resolved phonon properties to macroscopic heat transport in anisotropic crystals.

The cubic mean phase velocity of the acoustic modes measured by BLS (Figures~\ref{fig:Fig1}(d)-(f)) and calculated near $\Gamma$ (Figures~S2(a)-(c)) closely matches the accumulated cubic mean group velocity shown in Figures~S2(d)-(f) when only phonons with frequencies up to 3~THz are included. This agreement is expected because this frequency range is dominated by acoustic modes near $\Gamma$, for which group and phase velocities coincide. At higher frequencies, the angular dependence changes substantially. Increasing the frequency cutoff from 3 to 4~THz introduces a distinct anisotropy arising from acoustic modes near the Brillouin-zone boundary and the lowest optical branches, as shown by the calculated phonon dispersion relation in section S7 of the SI. Including higher-frequency optical modes further enhances this anisotropy. The phonon MFPs (Figures~S2(g)-(i)) exhibit a similar but weaker evolution, indicating that their directional variation is less pronounced than that of the group velocities. These results show that the anisotropy of low-frequency acoustic velocities alone cannot account for the measured thermal-conductivity anisotropy. We therefore examine the frequency-resolved contributions to the thermal conductivity.

Figures~\ref{fig:Fig1}(g)-(i) show the resulting frequency-accumulated, angle-resolved thermal conductivities. These quantities are obtained by decomposing the \textit{ab initio} Green--Kubo thermal conductivity into mode-resolved contributions, including both particle-like and wave-like transport, and summing contributions from modes below the specified frequency cutoff~\cite{aiGK,Knoop2023_PRB,Zhao2025}. The angular anisotropy of the macroscopic thermal conductivity measured by ATT is recovered only when phonon contributions above 4~THz are included. In contrast, when the accumulation is restricted to acoustic modes up to  3~THz (green lines in Figures~\ref{fig:Fig1}(g)-(i)), the resulting angular patterns closely resemble those of the cubic mean acoustic phase velocity, $\bar{v}$ (green lines in Figures~\ref{fig:Fig1}(d)-(f)). Thus, low-frequency acoustic phonons alone do not reproduce the bulk thermal-conductivity anisotropy. The distinct angular dependence of higher-frequency mode contributions is therefore essential for establishing the macroscopic anisotropy pattern.

\begin{figure*}[h!]
    \centering
    \includegraphics[width=1\linewidth]{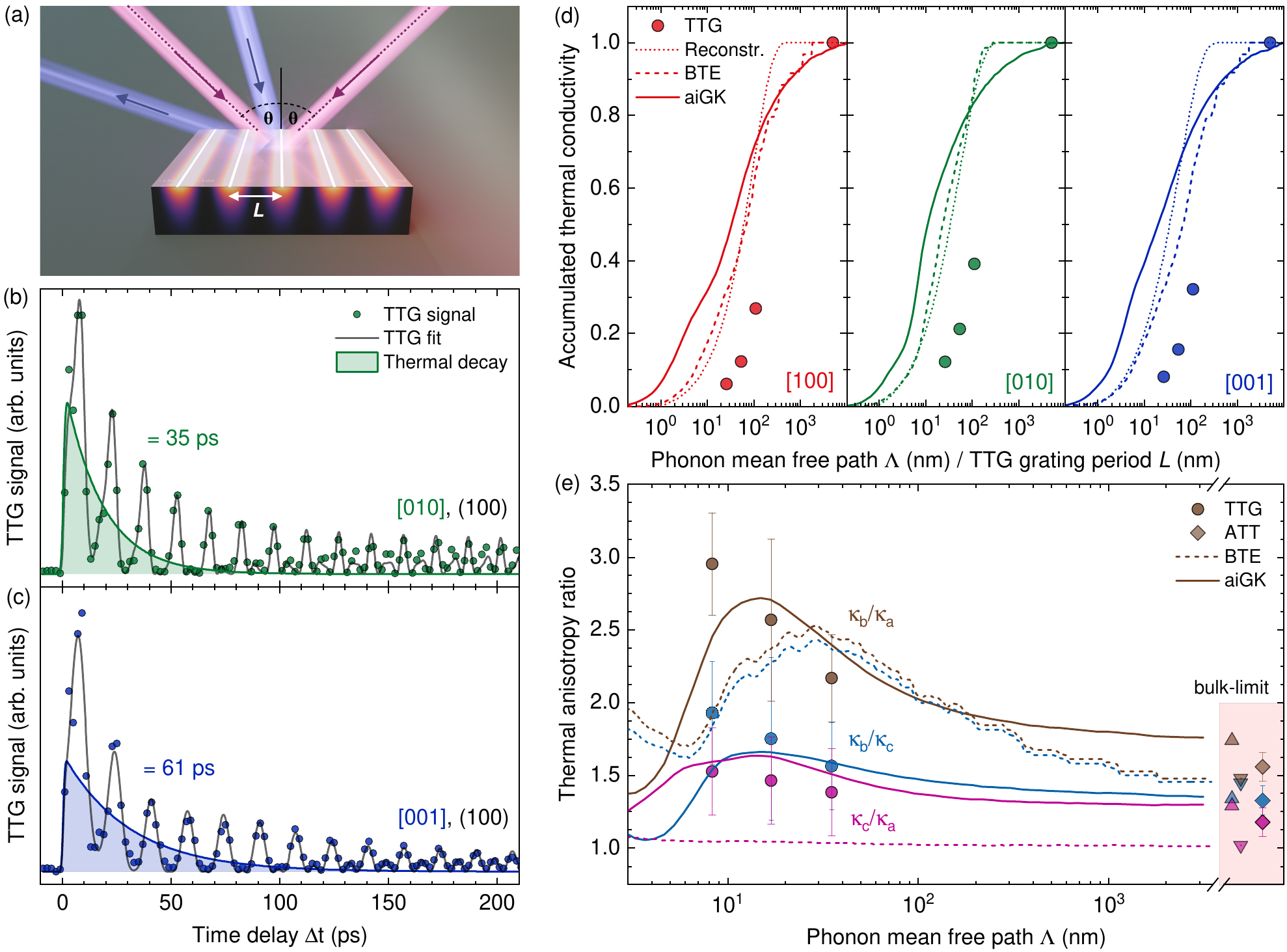}
    \caption{(a) Schematic illustration of the transient thermal grating (TTG) generated and probed by time-delayed femtosecond EUV pulses, (b)-(c) Representative time traces of EUV-TTG signal on the (100) plane of \betagao{} along the [001] and [010] directions. The exponential decay shown in blue accounts for the in-plane thermal decay perpendicular to the grating lines. (d) Normalized effective thermal conductivities measured by TTG as a function of grating spacing (circles), together with reconstructed (dotted lines), BTE (dashed lines), and aiGK (solid lines) accumulated thermal conductivities as a function of phonon MFP. The reconstructed curves are obtained from the measured grating-spacing-dependent conductivities (see main text). All datasets are normalized to their corresponding bulk thermal conductivities listed in Table~\ref{thermal_table}. (e) Thermal-anisotropy ratios as a function of phonon MFP. Circles denote TTG-derived experimental ratios displayed on the model-derived phonon-MFP scale obtained from the MFP-reconstruction. Diamonds denote experimental ATT bulk-limit values, and triangles denote calculated bulk-limit values. Neither set of bulk-limit reference points is assigned a finite grating spacing or phonon MFP. Solid and dashed lines represent aiGK and BTE calculations, respectively.}
    \label{fig:Fig2}
\end{figure*}

The distinct anisotropies of the low-frequency acoustic and higher-frequency phonons raise the question of how the effective thermal anisotropy evolves when different phonon populations are sampled over nanoscale transport lengths. To address this question experimentally, we conducted transient thermal grating (TTG) measurements with variable grating spacings, allowing us to selectively probe phonon contributions with different characteristic mean free paths (MFPs). In TTG, two pulsed laser beams interfere at the sample surface, creating a periodic intensity pattern and thereby inducing a ``thermal grating'', \textit{i.e.}, a spatially harmonic temperature modulation. Figure~\ref{fig:Fig2}(a) displays a schematic of the experimental configuration. The relaxation of the thermal grating, monitored via diffraction of a third laser beam, yields the thermal diffusivity along the in-plane direction perpendicular to the grating lines.

As the grating spacing \(L\) becomes comparable to phonon MFPs, the contribution of phonons with MFPs longer than \(L\) is progressively suppressed~\cite{maznev_onset_2011, johnson_direct_2013}. By decreasing \(L\), the relative contribution of shorter-MFP phonons is enhanced. Tuning \(L\) therefore enables a systematic exploration of the length-scale-dependent phonon contributions to thermal transport. However, accessing length scales relevant to short-MFP high-frequency phonons is not possible with visible-light TTG because diffraction limits the spatial resolution. This limitation can be overcome by utilizing EUV light at the TIMER instrument of the FERMI free-electron laser, which enables grating spacings as small as 28~nm in the present experiments and thereby accesses the length-scale regime where short-MFP phonons contribute substantially to thermal transport.

Figures~\ref{fig:Fig2}(b)-(c) show two representative EUV-TTG waveforms measured for the [001] and [010] directions of a (100) sample. The transients exhibit several oscillations with distinct frequencies and relatively long relaxation times, which are attributed to surface acoustic waves (SAWs). An exponential decay (shaded region) is evident, corresponding to the relaxation time of the thermal grating. For the quantitative evaluation of the signal, the waveforms are fitted using a model that combines the exponential decay of the thermal grating with the oscillations of the SAWs~\cite{foglia_extreme_2023, fainozzi_stimulated_2024} (cf. Materials and Methods). Similar transients were obtained at various grating spacings and on the (001) sample. The in-plane thermal diffusivity along the grating wavevector, i.e., perpendicular to the grating lines, is obtained as $\alpha=L^2/4\pi^2\tau_{th}$ (see also Materials and Methods). 

By measuring the dependence of the thermal conductivity on characteristic length scales through variation of the grating period in TTG experiments, it is possible to reconstruct the accumulated thermal conductivity as a function of phonon MFP. This methodology is commonly referred to as the reconstruction method, originally proposed by Minnich~\cite{Minnich2012PRL}, and subsequently refined by other researchers~\cite{Hua2015JAP, Cuffe2015PRB, Zeng2019JAP}. In this framework, the grating-period-dependent effective thermal conductivity along direction \(i\), \(\kappa_i(L)\), is expressed as:
$
k_i = \int S(x_i)\, f(\Lambda_\omega) \, d\Lambda_\omega,
$
where $f(\Lambda_\omega)$ is the differential MFP distribution function, $S(x_i)$ is the suppression function, and $x_i = \Lambda_\omega/L$ is the Knudsen number. The suppression function \(S(x_i)\) quantifies the extent to which phonon modes are attenuated by finite-length-scale effects and thus governs the deviation from purely diffusive transport.

The EUV-TTG experiments allow us to obtain three independent measurements of the thermal conductivity $k_{i}(L)$ for each principal axis, each corresponding to a different grating period $L$. As $L$ becomes comparable to or smaller than the phonon MFPs, the contribution of long-MFP phonons to the measured thermal conductivity is increasingly suppressed. This suppression is described by a geometry-specific function $S(\Lambda_\omega, L)$ that depends on both the phonon MFP and the system length scale. To obtain the accumulated thermal conductivity function, we integrate the differential expression as: $F(\Lambda_\omega) = \int_0^{\Lambda_\omega} f(\Lambda') \, d\Lambda'$. The corresponding inverse problem was solved using a convex optimization framework. Within the assumptions of the suppression model, $F$ is an intrinsic material property, independent of the particular experimental conditions used to probe it.

Figure~\ref{fig:Fig2}(d) compares the reconstructed accumulated thermal conductivities (dotted lines) obtained from the TTG data with the BTE and aiGK results (dashed and solid lines, respectively). The accumulated conductivity curves represent the fractional contribution to the total thermal conductivity from phonons with MFPs shorter than a given value and therefore characterize the MFP-resolved accumulation of thermal transport. For comparison, the figure also includes the effective thermal diffusivities obtained from the TTG experiments, normalized to their corresponding bulk values, for all crystallographic directions and grating spacings. Note that the horizontal axis has different meanings for the two datasets. The TTG data are plotted as a function of the grating spacing \(L\), whereas the reconstructed and calculated accumulation curves are plotted as a function of phonon MFP. A TTG measurement at a given \(L\) reflects the MFP distribution weighted by the corresponding TTG suppression function rather than the accumulated thermal conductivity at \(\Lambda=L\).

At short MFPs, the reconstructed accumulation functions closely follow the BTE calculations, whereas the aiGK curves exhibit a more gradual initial accumulation. At longer MFPs, the aiGK accumulation remains close to or below the BTE result, consistent with the additional anharmonic scattering processes captured beyond the lowest-order perturbative treatment. Despite these quantitative differences in the MFP-resolved distributions, the reconstructed and calculated curves exhibit the same directional trends and characteristic MFP range over which the thermal conductivity accumulates. This agreement supports the interpretation that the measured length-scale dependence arises from the progressive suppression of long-MFP phonon contributions.

Figure~\ref{fig:Fig2}(e) compares the anisotropy ratios measured by TTG with those calculated from the accumulated MFP-resolved thermal conductivities by BTE and aiGK. To place these datasets on a common axis, the experimental TTG ratios are displayed on an effective MFP scale derived from the reconstruction. This scale represents the characteristic MFP range sampled at each grating period within the assumed suppression model and does not imply a one-to-one correspondence between grating period and phonon MFP. The calculated ratios represent the anisotropy of the accumulated thermal conductivities contributed by all phonon modes with MFPs below the respective cutoff. For reference, the experimental ATT values and the calculated bulk values are included as bulk-limit points. They represent the macroscopic diffusive limit and are therefore not assigned a physical TTG grating period or an effective MFP coordinate. The experimental and calculated anisotropy ratios exhibit the same plane-dependent evolution, with the strongest increase toward short effective MFPs in the (001) plane.

The data reveal a progressive but strongly plane-dependent increase in in-plane thermal anisotropy as the grating spacing decreases. The strongest increase occurs in the (001) plane, where the anisotropy nearly doubles relative to its macroscopic value. A more moderate enhancement is observed in the (100) plane, whereas the anisotropy in the (010) plane varies only weakly over the investigated range. Thus, reducing the characteristic transport length scale does not simply suppress the thermal conductivity, but selectively changes the directional contrast of the phonon populations contributing to heat transport. The enhanced anisotropy at short grating periods indicates that suppressing long-MFP contributions increases the relative weight of phonons with stronger directional contrast. Together with the frequency-resolved aiGK results and the MFP-dependent BTE and aiGK calculations, this behavior identifies higher-frequency, shorter-MFP phonons as the primary origin of the enhanced nanoscale anisotropy. Conversely, longer-MFP, predominantly lower-frequency phonons attenuate the directional contrast and reduce the accumulated anisotropy toward its macroscopic value. The weak length-scale dependence in the (010) plane is consistent with the frequency-resolved results in Figure~\ref{fig:Fig1}. In this plane, the higher-frequency phonons exhibit only weak anisotropy compared to the more pronounced anisotropy in the two other planes. Consequently, suppressing lower-frequency, longer-MFP contributions produces only a small change in the accumulated in-plane thermal anisotropy. 

Taken together, the TTG measurements and MFP-resolved calculations show that the thermal anisotropy of \betagao{} is not fully described by a single bulk conductivity ratio. Instead, it depends on the phonon population sampled by the thermal field. At nanoscale grating periods, the reduced contribution of long-MFP phonons reveals the stronger directional contrast carried by shorter-MFP phonons, whereas the macroscopic response reflects the combined contribution of phonon populations with distinct anisotropies.

In summary, we establish the nanoscale phononic origin of thermal anisotropy in \(\beta\)-Ga\(_2\)O\(_3\) by combining anisotropic thermoreflectance thermometry (ATT), angle-resolved Brillouin light scattering (BLS), and extreme UV transient thermal grating (EUV-TTG) spectroscopy with \textit{ab initio} Green-Kubo and Boltzmann transport equation calculations. This complementary approach connects the macroscopic thermal-conductivity tensor with the acoustic, frequency-dependent, and length-scale-dependent properties of heat-carrying phonons.

The weak anisotropy of low-frequency acoustic velocities measured by BLS is insufficient to explain the measured thermal-conductivity anisotropy. Instead, frequency-resolved calculations show that the bulk anisotropy emerges only when phonon modes above approximately \(4\)~THz are included, whereas contributions from acoustic modes below \(3\)~THz remain comparatively weakly anisotropic. Consistently, EUV-TTG measurements reveal a plane-dependent enhancement of the in-plane thermal anisotropy as the grating period is reduced to 28~nm, reaching nearly twice the macroscopic value in the (001) plane and showing progressively weaker increases in the (100) and (010) planes.

These results identify high-frequency, short-MFP phonons as the dominant origin of the pronounced thermal anisotropy in \(\beta\)-Ga\(_2\)O\(_3\), while long-MFP, low-frequency phonons attenuate this directional contrast at macroscopic length scales. More broadly, they show that thermal anisotropy in low-symmetry crystals can change substantially when heat flow is confined to the nanoscale regime, providing a route to resolve and engineer nanoscale heat flow for thermal management in ultra-wide-bandgap power electronics.

\section*{Supporting Information}
The Supporting Information contains additional details on the symmetry-adapted coordinate system and crystallographic directions (Section S1), the elastic tensor and Young’s moduli (Section S2), and phonon velocities and mean free paths from \textit{ab initio} Green--Kubo calculations (Section S3). It further describes angular-resolved in-plane thermal-conductivity measurements by anisotropic thermoreflectance thermometry (Section S4), in-plane thermal-diffusivity measurements by beam-offset frequency-domain thermoreflectance (Section S5), thermal anisotropy in doped crystals (Section S6), and the phonon dispersion (Section S7). Section S8 provides an overview of thermal conductivities reported in the literature.

\section*{Acknowledgments}
We acknowledge funding by the Deutsche Forschungsgemeinschaft (DFG, German Research Foundation) under project number 446185170 and the Spanish Ministerio de Ciencia e Innovación (MICINN) for its support through Grant CEX2023-001263-S and CEX2021-001214-S in the framework of the Spanish Severo Ochoa Centre of Excellence program, and Grants PID2020-119777GB-I00 (THERM2MAIN),  PDC2021-121814-I00 (COVEQ), PDC2023-145934-I00 (COOL), and PID2024-162811NB-I00 (TAPAS). We also acknowledge Grant 2021-SGR-004 4 4 (NANOPTO) from the Catalan agency AGAUR. This work was partially funded by the ERC Advanced Grant TEC1p (European Research Council, Grant Agreement No.740233). The research was supported by the project Laserlab-Europe under Grant Agreement 871124 from the EU Framework Programme for Research and Innovation HORIZON 2020. We thank the Centro de Supercomputaci\'on de Galicia (CESGA) for the use of their computational resources. We thank Kingsley Egbo for supporting atomic force microscopy measurements. We thank Roberts Blukis from Leibniz-Institut für Kristallzüchtung for critical reading of the paper. The authors acknowledge the use of Perplexity (accessed October 2026) to assist with language editing, check numerical consistency, and identify potential inconsistencies in the scientific discussion. All AI-generated suggestions were independently evaluated by the authors, and any resulting revisions were reviewed and approved by the authors, who take full responsibility for the accuracy and integrity of the work.

\clearpage

\section{Materials and methods}
\subsection{Samples}
Ga$_{2}$O$_{3}$ crystal samples with a size of \mbox{$5 \times 5 \times 0.5$~mm$^{3}$} used in the present study were prepared from a nominally undoped, two-inch-diameter single crystal grown by the Czochralski method utilizing an Ir crucible and oxidizing growth atmosphere, as described in detail elsewhere \cite{Galazka2017,Galazka2021}. The single crystals exhibit high structural quality, with no twins, low-angle grain boundaries, or nanopipes, while the dislocation density did not exceed values of \mbox{$\sim 10^{3}$ cm$^{-2}$ \cite{Galazka2022}.} The investigated (100)-, (010)-, and (001)-oriented \mbox{$\beta-$Ga$_{2}$O$_{3}$} crystal planes received a chemical-mechanical polishing (CMP) resulting in low RMS surface roughness of smaller than 0.2~nm. The samples are semiconducting with the free electron concentration of \mbox{$3.4\cdot 10^{17}$ cm$^{-3}$} and an electron mobility of \mbox{118 cm$^{2}$V$^{-1}$s$^{-1}$} according to Hall effect measurements \cite{Irmscher2011}. 

\subsection{Anisotropic thermoreflectance thermometry} 
The anisotropic thermoreflectance thermometry (ATT) method \cite{Perez2022} is a contactless optical technique to accurately measure the thermal conductivity tensor of isotropic and anisotropic materials. The experimental setup is based on a pump-probe configuration. A 405~nm pump laser is used to induce time-periodic heating of the sample, while a 532~nm probe laser monitors the thermal response by detecting changes in optical reflectivity, which are proportional to the surface temperature oscillations.

A key feature of the method is the generation of a line-shaped heat source. This is achieved by passing the pump and probe laser beams through a custom-designed diffractive optical element (DOE), which converts the original Gaussian intensity distribution of the lasers into a narrow, elongated line with uniform intensity along its length and a Gaussian distribution across its width. This geometry simplifies thermal analysis by eliminating temperature gradients along the line direction, thus, in-plane heat flow is limited to the direction perpendicular to the line heater, making it particularly well-suited for anisotropic systems.
The sample is mounted on a rotational stage, allowing precise alignment with the line heater and systematic measurements of thermal conductivity along various in-plane crystallographic directions. By analyzing the temperature response as a function of the logarithm of the modulation frequency of the pump laser, the thermal conductivity tensor components are extracted using the slope method. This method calculates thermal conductivity directly from the slope of the linear relationship between the amplitude of the temperature oscillation and the logarithmic frequency, significantly reducing the need for complex numerical modeling.
This approach accommodates diverse sample geometries, including bulk materials, thin films, and multilayer systems, with mathematical solutions tailored for each case. The experimental accuracy of the ATT method is approximately 5\%, with a temperature sensitivity of $\approx 3$~mK. The contactless nature of the technique minimizes sample preparation and avoids issues associated with electrical contacts, making it ideal for studying a wide range of materials, from isotropic systems such as silicon and germanium to anisotropic ones such as \betagao{} and Kapton tape.

\subsection{Extreme UV transient thermal grating}
The transient thermal grating (TTG) technique is conventionally implemented using tabletop optical lasers. To achieve nanoscale sensitivity, the TIMER beamline at FERMI \cite{Mincigrucci2018NIMA} uses extreme ultraviolet (EUV) pulses for both exciting and probing the TTG. This technique generates a transient grating by the interference of two ultrafast EUV pump beams on the sample surface, creating a periodic heating pattern with spatial periodicity defined as
\begin{equation}
L = \frac{\lambda}{2\sin\theta},
\end{equation}
where \(\lambda\) is the pump wavelength and \(\theta\) is the half-crossing angle of the beams. The transient grating serves as a nanoscale heat source, resulting in a sinusoidal temperature profile (thermal grating) at the sample surface. The subsequent thermal expansion launches counter-propagating SAWs at the TTG wavelength {L}. Both the thermal grating and the propagating SAWs modulate the surface profile. The dynamics of such modulations can be probed by a time-delayed EUV pulse, which is diffracted by the induced TTG in reflection geometry (see Fig. 2(a)), providing time-resolved data on the dynamics of heat and elastic wave propagation. Measurements were carried out at the TIMER instrument, using $\theta = 13.8^\circ$ and $\lambda = 13.3$, $26.6$, and $53.2$~nm for the pump beams. The wavelength of the probe beam was $13.3$~nm, the estimated pulse duration was about 40~fs (full-width-at-half maximum), and the repetition rate of the FEL was 50~Hz. 

By analyzing the decay of the TTG signal, the thermal diffusivity $\alpha$ of the material can be determined by the relationship:
\begin{equation}
\alpha = \frac{L^2}{4 \pi^2 \tau_{\text{th}}},
\end{equation}
where \(\tau_{\mathrm{th}}\) is obtained by fitting the slow exponential decrease in the diffraction signal.

The TTG transients were fitted using the following expression~\cite{fainozzi_stimulated_2024,maznev_generation_2021}, including one SAW oscillation:\begin{equation}
I(\Delta t) = \Biggl| \frac{1}{2} \Bigl[1 + \operatorname{erf}\Bigl(\frac{\Delta t}{\sigma}\Bigr)\Bigr] 
\Bigl( A \exp\Bigl(-\frac{\Delta t}{\tau_{\mathrm{th}}}\Bigr) 
- A_{\mathrm{SAW}} \sin\bigl(2\pi \nu_{\mathrm{SAW}} \Delta t + \varphi_{\mathrm{SAW}}\bigr) \Bigr) \Biggr|^2,
\end{equation}
where $\sigma$ is the TTG signal rise, $\nu_{SAW}$ and $\varphi_{\mathrm{SAW}}$ are the SAW frequency and phase, respectively, and $A$ and $A_{SAW}$ are scaling amplitudes.

\subsection{Brillouin light scattering} 
Brillouin light scattering (BLS) experiments were conducted at room temperature using a high-contrast tandem Fabry-Perot interferometer (J.R. Sandercock Scientific Instruments) combined with a single-mode continuous-wave (CW) laser operating at a wavelength of \(\lambda_0 = 532\)~nm. The experiments utilized the transmission (90°) BLS scattering geometry in s–s polarization of the incident and scattered light, a configuration specifically chosen to mitigate issues related to unknown optical birefringence. This arrangement eliminates the need for prior knowledge of the refractive index, allowing the magnitude of the acoustic wave vector, \(q\), to be directly assigned based on the geometry. In this setup, \(q\) lies entirely within the plane of the sample and its magnitude is given by \(q = 2\sqrt{2}\pi / \lambda_0\). By rotating the crystal around its surface normal, the direction of \(q\) relative to the crystallographic orientation of the sample can be precisely controlled.

The phase velocity of the acoustic phonons, \(v\), is calculated as \(v = 2\pi f / q\), where \(f\) is the Brillouin shift obtained from the experimental spectra. This methodology was applied to three specific crystallographic orientations of the \(\beta\)-Ga\(_2\)O\(_3\) crystal: the a-, b-, and c-planes, corresponding to the (100), (010), and (001) planes, respectively. For each orientation, the angular dispersion of the acoustic phonon phase velocity, \(v(\varphi)\), was measured, enabling a comprehensive analysis of the anisotropic acoustic properties of \(\beta\)-Ga\(_2\)O\(_3\). This approach provides detailed information on the phononic behavior of the material as a function of its crystallographic orientation.

\subsection{Ab initio Green-Kubo Calculations}
\label{sec:aiGK}
The \textit{ab initio} Green-Kubo method~(aiGK)~\cite{aiGK,Knoop2023_PRB} is used for calculating the anisotropic thermal conductivity using equilibrium \textit{ab initio} molecular dynamics~(aiMD) simulations. By this means, anharmonic contributions both to particle- and wave-like transport~\cite{Simoncelli.2022} are captured to all orders. Compared to perturbative approaches such as the BTE, cf.~Sec.~\ref{sec:BTE}, in which only the lowest order of anharmonicity is typically accounted for, the aiGK method hence provides more accurate predictions, especially for strongly anharmonic, thermally insulating systems~\cite{Knoop2023_PRL}. In practice,  the thermal conductivity is obtained from the auto-correlation of the heat-flux along aiMD trajectories. To this end, the conductive (often called virial) contributions of the heat flux are calculated~\cite{aiGK} from the analytic atomic stress tensor~\cite{stresses} from first principles. See Ref.~\cite{Knoop2023_PRB} for more details on the method and its implementation in {\tt FHI-vibes}~\cite{fhi-vibes}.

The underlying \textit{ab initio} calculations were carried out with the all-electron \textit{ab initio} simulation code {\tt FHI-aims}~\cite{fhi-aims}, which utilizes numeric atomic orbitals as a basis set. For the presented \betagao{} results, we used the Perdew-Burke-Ernzerhof (PBE) exchange-correlation functional~\cite{PerdewPRL96}, ``light'' basis sets and numerical settings as provided by \texttt{FHI-aims}~\cite{fhi-aims,Carbogno.2022}, and a $k$-point density of $2\,\mathrm{kpt}/\mathring{\mathrm{A}}^{-1}$. The aiMD simulations~(timestep 5~fs) were run in an extended $160$-atom supercell, generated from the primitive cell via the transformation matrix $[[1, -1, 0],[4, 4, 0],[0, 0, 2]]$, using {\tt FHI-vibes}~\cite{fhi-vibes} to orchestrate the calculations and post-process the simulations. In particular, we followed the workflow for noise reduction and finite-size corrections described in Ref.~\cite{Knoop2023_PRB} for the aiGK calculations. To capture thermal expansion and all anharmonic contributions to it, the lattice parameters at finite temperature were determined by minimizing the expectation value of the external pressure observed in NVT aiMD simulations using the Langevin thermostat. Uncorrelated snapshots from the NVT simulations at 300~K were used as initial configurations for ten independent aiGK calculations performed in the NVE ensemble. Each of the trajectories has a length of $120\,\mathrm{ps}$ and the heat flux is calculated each second time step. 

Harmonic force constants are extracted from the aiMD simulations~\cite{Esfarjani.2008} using the temperature-dependent effective potentials~\texttt{TDEP}~\cite{TDEP2nd, TDEP2024} code. These force constants are also used to decompose the thermal conductivity $\kappa^{\alpha\beta}$ into the contributions $\kappa^{\alpha\beta}(\vec{q},j)$ of the individual modes with branch index $j$ and wave vector $\vec{q}$, using the approach proposed in Ref.~\cite{McGaughey.2004}. A detailed description of the implementation can be found in Ref.~\cite{Knoop2023_PRB} and Ref.~\cite{Zhao2025}. The decomposition also yields the relaxation times $\tau(\vec{q},j)$, which are obtained from the decay of the mode-resolved heat-flux autocorrelation functions. The accumulated thermal conductivity as a function of the phonon frequency, shown in Fig.~\ref{fig:Fig1}(g)-(i), is calculated via
\begin{equation}
\kappa^{\alpha\beta}_{\mathrm{accum}}(\omega_0)=
\sum_{\vec{q},j}\kappa^{\alpha\beta}(\vec{q},j)\;
\Theta\bigl(\omega_0-\omega(\vec{q},j)\bigr),
\label{eq:accum_freq}
\end{equation}
with $\omega(\vec{q},j)$ being the frequency of branch $j$ at the wave vector $\vec{q}$, $\omega_0$ being the cutoff frequency, and $\Theta$ being the Heaviside step function. The accumulated thermal conductivity as a function of the phonon MFP, shown in Fig.~\ref{fig:Fig2}(d),(e), is calculated via
\begin{equation}
\kappa^{\alpha\beta}_{\mathrm{accum}}(\Lambda_0,\mathbf{n})=
\sum_{\vec{q},j}\kappa^{\alpha\beta}(\vec{q},j)\;
\Theta\bigl(\Lambda_0-|\vec{\Lambda}(\vec{q},j)\cdot
\mathbf{n}|\bigr),
\label{eq:accum_mfp}
\end{equation}
with $\vec{\Lambda}(\vec{q},j)=\vec{v}_g(\vec{q},j)\,\tau(\vec{q},j)$ being the MFP of branch $j$ at the wave vector $\vec{q}$, $\mathbf{n}$ being the transport direction, and
$\Lambda_0$ being the cutoff MFP. In both cases, the conductivity along $\mathbf{n}$ is obtained from the accumulated tensor via the same quadratic form as $\kappa_{\mathrm{equiv}}(\mathbf{n})$ in the main text, i.e., $\mathbf{n}^{\mathrm{T}}\boldsymbol{\kappa}_{\mathrm{accum}}\mathbf{n}$. The accumulations are scaled such that the sum over all modes reproduces the bulk aiGK tensor given in Table~\ref{thermal_table}.

\subsection{Boltzmann Transport Equation }
\label{sec:BTE}
We calculated the second- and third-order interatomic force constants (IFCs) within density-functional theory (DFT), using the VASP code~\cite{KressePRB96} with the generalized gradient approximation (GGA) in the parameterization due to Perdew, Burke, and Ernzerhof~\cite{PerdewPRL96} for the exchange-correlation energy functional and a plane wave cutoff of 400~eV with the projector augmented-wave method~\cite{KressePRB99}. The harmonic IFCs were calculated from finite differences in a $3 \times 3 \times 3$ supercell, while we used a $2 \times 2 \times 2$ supercell for the anharmonic ones, limiting the interactions to eighth neighbors. The inequivalent displacements needed to obtain the IFCs were obtained with the phonopy~\cite{TogoSM15} and thirdorder.py~\cite{LiCPC14} codes. We iterated the electronic self-consistency loop until changes in the total energy and eigenvalues were lower than $10^{-7}$~eV.
Once the IFCs are obtained, we solve iteratively the linearized Boltzmann Transport Equation (BTE) for phonons with the almaBTE code~\cite{CarreteCPC17} on a $14 \times 14 \times 14$, {\bf q}-point grid, obtaining lattice thermal conductivity as
\begin{equation}
\kappa_l^{\alpha\beta} = \frac{1}{k_B T^2 \Omega N} \sum_\lambda f_0(f_0+1)(\hbar \omega_\lambda)^2 v_{\alpha,\lambda} F_{\beta,\lambda}
\end{equation}
where $\alpha$ and $\beta$ are the three coordinate directions $x$, $y$, and $z$; and $k_B$, $T$, $\Omega$, and $N$ are the Boltzmann constant, the temperature, the volume of the unit cell and the number of q-points, respectively. $F_{\beta,\lambda}$ is initially taken to be equal to $\tau_{\lambda}v_{\beta,\lambda}$, where $\tau_{\lambda}$ is the lifetime of the phonon mode $\lambda$ within the relaxation time approximation (RTA). Starting from this initial guess, the solution is then obtained iteratively and $F_{\beta,\lambda}$ takes the general form $\tau_{\lambda}v_{\beta,\lambda}+\Delta_{\beta,\lambda}$. Scattering from isotopic disorder, besides anharmonic three-phonon scattering, is also included considering the natural isotopic distributions of Ga and O through the model due to Tamura~\cite{tamura_isotope_1983}.

Our calculations are similar to those of Santia et al. \cite{Santia2015}, though we used a larger supercell for harmonic interatomic force constants ($3 \times 3 \times 3$ vs. $2 \times 2 \times 2$), a larger neighbor cutoff for the anharmonic interatomic force constants (eighth neighbors vs. third neighbors), and a thicker {\bf q}-point grid to solve the BTE ($14 \times 14 \times 14$ vs. $10 \times 10 \times 10$). In fact, we have observed that this refinement is important to achieve converged values of the thermal conductivity, which are also in very good agreement with our experimental data.

The \betagao{} crystal is monoclinic and its conventional unit cell contains 20 atoms (space-group C2/m). We adopt the scheme proposed by H. Peelaers and C. G. Van de Walle\cite{Peelaers2015} to reduce the 20-atom conventional unit cell to a 10-atom primitive cell. We obtain the following structural parameters, referred to the conventional unit cell: a=12.2656~Å, b=3.0575~Å, c=5.8256~Å, and $\beta=103.64^\circ$. 

\bibliography{bibliography-betagao}

@article{passler_hyperbolic_2022,
	title = {Hyperbolic shear polaritons in low-symmetry crystals},
	volume = {602},
	copyright = {2022 The Author(s)},
	issn = {1476-4687},
	url = {https://www.nature.com/articles/s41586-021-04328-y},
	doi = {10.1038/s41586-021-04328-y},
	language = {en},
	number = {7898},
	urldate = {2024-05-15},
	journal = {Nature},
	author = {Passler, Nikolai C. and Ni, Xiang and Hu, Guangwei and Matson, Joseph R. and Carini, Giulia and Wolf, Martin and Schubert, Mathias and Alù, Andrea and Caldwell, Joshua D. and Folland, Thomas G. and Paarmann, Alexander},
	month = feb,
	year = {2022},
	note = {Publisher: Nature Publishing Group},
	pages = {595--600},
}

@article{schubert_anisotropy_2016,
	title = {Anisotropy, phonon modes, and free charge carrier parameters in monoclinic \${\textbackslash}ensuremath\{{\textbackslash}beta\}\$-gallium oxide single crystals},
	volume = {93},
	url = {https://link.aps.org/doi/10.1103/PhysRevB.93.125209},
	doi = {10.1103/PhysRevB.93.125209},
	number = {12},
	urldate = {2024-05-15},
	journal = {Physical Review B},
	author = {Schubert, M. and Korlacki, R. and Knight, S. and Hofmann, T. and Schöche, S. and Darakchieva, V. and Janzén, E. and Monemar, B. and Gogova, D. and Thieu, Q.-T. and Togashi, R. and Murakami, H. and Kumagai, Y. and Goto, K. and Kuramata, A. and Yamakoshi, S. and Higashiwaki, M.},
	month = mar,
	year = {2016},
	note = {Publisher: American Physical Society},
	pages = {125209},
}

@article{basov_towards_2017,
	title = {Towards properties on demand in quantum materials},
	volume = {16},
	copyright = {2017 Springer Nature Limited},
	issn = {1476-4660},
	url = {https://www.nature.com/articles/nmat5017},
	doi = {10.1038/nmat5017},
	language = {en},
	number = {11},
	urldate = {2024-05-15},
	journal = {Nature Materials},
	author = {Basov, D. N. and Averitt, R. D. and Hsieh, D.},
	month = nov,
	year = {2017},
	note = {Publisher: Nature Publishing Group},
	pages = {1077--1088},
}

@article{Cho2021,
author = {Cho, Jeong Bin and Jung, Gunwoo and Kim, Kyuheon and Kim, Jihun and Hong, Soon Ku and Song, Jung Hoon and Jang, Joon Ik},
doi = {10.1021/acs.jpcc.0c08413},
issn = {19327455},
journal = {J. Phys. Chem. C},
number = {2},
pages = {1432--1440},
title = {{Highly Asymmetric Optical Properties of $\beta-\textrm{Ga}_{2}\textrm{O}_{3}$ as Probed by Linear and Nonlinear Optical Excitation Spectroscopy}},
volume = {125},
year = {2021},
url = {https://doi.org/10.1021/acs.jpcc.0c08413}
}

@article{Furthmueller2016,
author = {Furthm{\"{u}}ller, J. and Bechstedt, F.},
doi = {10.1103/PhysRevB.93.115204},
issn = {2469-9950},
journal = {Phys. Rev. B},
month = {mar},
number = {11},
pages = {115204},
title = {Quasiparticle bands and spectra of $\textrm{Ga}_{2}\textrm{O}_{3}$ polymorphs},
url = {https://link.aps.org/doi/10.1103/PhysRevB.93.115204},
volume = {93},
year = {2016}
}

@article{Galazka2014,
author = {Galazka, Zbigniew and Irmscher, Klaus and Uecker, Reinhard and Bertram, Rainer and Pietsch, Mike and Kwasniewski, Albert and Naumann, Martin and Schulz, Tobias and Schewski, Robert and Klimm, Detlef and Bickermann, Matthias},
doi = {10.1016/j.jcrysgro.2014.07.021},
issn = {00220248},
journal = {J. Cryst. Growth},
month = {oct},
pages = {184--191},
publisher = {Elsevier},
title = {On the bulk $\beta-\mathrm{Ga}_{2}\mathrm{O}_{3}$ single crystals grown by the Czochralski method},
doi = {10.1016/j.jcrysgro.2014.07.021},
url = {https://linkinghub.elsevier.com/retrieve/pii/S0022024814005193},
volume = {404},
year = {2014}
}

@article{Galazka2017,
author = {Galazka, Zbigniew and Uecker, Reinhard and Klimm, Detlef and Irmscher, Klaus and Naumann, Martin and Pietsch, Mike and Kwasniewski, Albert and Bertram, Rainer and Ganschow, Steffen and Bickermann, Matthias},
doi = {10.1149/2.0021702JSS/XML},
issn = {2162-8769},
journal = {ECS J. Solid State Sci. Technol.},
month = {sep},
number = {2},
pages = {Q3007--Q3011},
publisher = {The Electrochemical Society},
title = {{ Scaling-Up of Bulk $\beta-\mathrm{Ga}_{2}\mathrm{O}_{3}$ Single Crystals by the Czochralski Method }},
url = {https://iopscience.iop.org/article/10.1149/2.0021702jss},
volume = {6},
year = {2017}
}

@article{Galazka2021,
author = {Galazka, Zbigniew and Ganschow, Steffen and Irmscher, Klaus and Klimm, Detlef and Albrecht, Martin and Schewski, Robert and Pietsch, Mike and Schulz, Tobias and Dittmar, Andrea and Kwasniewski, Albert and Grueneberg, Raimund and Anooz, Saud Bin and Popp, Andreas and Juda, Uta and Hanke, Isabelle M. and Schroeder, Thomas and Bickermann, Matthias},
doi = {10.1016/j.pcrysgrow.2020.100511},
issn = {09608974},
journal = {Prog. Cryst. Growth Charact. Mater.},
month = {feb},
number = {1},
pages = {100511},
publisher = {Elsevier Ltd},
title = {{Bulk single crystals of $\beta-\mathrm{Ga}_{2}\mathrm{O}_{3}$ and Ga-based spinels as ultra-wide bandgap transparent semiconducting oxides}},
url = {https://linkinghub.elsevier.com/retrieve/pii/S0960897420300383},
volume = {67},
year = {2021}
}

@article{Galazka2022,
author = {Galazka, Zbigniew},
doi = {10.1063/5.0076962},
issn = {0021-8979},
journal = {Journal of Applied Physics},
month = {jan},
number = {3},
pages = {031103},
publisher = {AIP Publishing LLC},
title = {{Growth of bulk $\beta-\mathrm{Ga}_{2}\mathrm{O}_{3}$ single crystals by the Czochralski method}},
url = {https://aip.scitation.org/doi/10.1063/5.0076962},
volume = {131},
year = {2022}
}

@article{Gopalan2020,
author = {Gopalan, Prashanth and Knight, Sean and Chanana, Ashish and Stokey, Megan and Ranga, Praneeth and Scarpulla, Michael A. and Krishnamoorthy, Sriram and Darakchieva, Vanya and Galazka, Zbigniew and Irmscher, Klaus and Fiedler, Andreas and Blair, Steve and Schubert, Mathias and Sensale-Rodriguez, Berardi},
doi = {10.1063/5.0031464},
issn = {00036951},
journal = {Appl. Phys. Lett.},
number = {25},
publisher = {AIP Publishing LLC},
title = {{The anisotropic quasi-static permittivity of single-crystal $\beta-\mathrm{Ga}_{2}\mathrm{O}_{3}$ measured by terahertz spectroscopy}},
volume = {117},
pages = {252103},
year = {2020},
url = {https://doi.org/10.1063/5.0031464}
}

@article{Graczykowski2015,
author = {Graczykowski, B. and Sledzinska, M. and Alzina, F. and Gomis-Bresco, J. and Reparaz, J. S. and Wagner, M. R. and {Sotomayor Torres}, C. M.},
doi = {10.1103/PhysRevB.91.075414},
issn = {1098-0121},
journal = {Physical Review B},
month = {feb},
number = {7},
pages = {075414},
title = {{Phonon dispersion in hypersonic two-dimensional phononic crystal membranes}},
url = {https://link.aps.org/doi/10.1103/PhysRevB.91.075414},
volume = {91},
year = {2015}
}

@article{Green2022,
author = {Green, Andrew J. and Speck, James and Xing, Grace and Moens, Peter and Allerstam, Fredrik and Gumaelius, Krister and Neyer, Thomas and Arias-Purdue, Andrea and Mehrotra, Vivek and Kuramata, Akito and Sasaki, Kohei and Watanabe, Shinya and Koshi, Kimiyoshi and Blevins, John and Bierwagen, Oliver and Krishnamoorthy, Sriram and Leedy, Kevin and Arehart, Aaron R. and Neal, Adam T. and Mou, Shin and Ringel, Steven A. and Kumar, Avinash and Sharma, Ankit and Ghosh, Krishnendu and Singisetti, Uttam and Li, Wenshen and Chabak, Kelson and Liddy, Kyle and Islam, Ahmad and Rajan, Siddharth and Graham, Samuel and Choi, Sukwon and Cheng, Zhe and Higashiwaki, Masataka},
doi = {10.1063/5.0060327},
issn = {2166-532X},
journal = {APL Materials},
month = {feb},
number = {2},
pages = {029201},
publisher = {AIP Publishing, LLC},
title = {{$\beta$-Gallium oxide power electronics}},
doi = {https://doi.org/10.1063/5.0060327},
url = {https://aip.scitation.org/doi/10.1063/5.0060327},
volume = {10},
year = {2022}
}

@article{Grundmann2017,
author = {Grundmann, Marius and Sturm, Chris and Kranert, Christian and Richter, Steffen and Schmidt-Grund, R{\"{u}}diger and Deparis, Christianne and Z{\'{u}}{\~{n}}iga-P{\'{e}}rez, Jes{\'{u}}s},
doi = {10.1002/pssr.201600295},
issn = {18626254},
journal = {Phys. Status Solidi Rapid Res. Lett.},
month = {jan},
number = {1},
pages = {1600295},
title = {{Optically anisotropic media: New approaches to the dielectric function, singular axes, microcavity modes and Raman scattering intensities}},
url = {https://onlinelibrary.wiley.com/doi/10.1002/pssr.201600295},
volume = {11},
year = {2017}
}

@article{Guo2015,
archivePrefix = {arXiv},
arxivId = {1412.7472},
author = {Guo, Zhi and Verma, Amit and Wu, Xufei and Sun, Fangyuan and Hickman, Austin and Masui, Takekazu and Kuramata, Akito and Higashiwaki, Masataka and Jena, Debdeep and Luo, Tengfei},
doi = {10.1063/1.4916078},
eprint = {1412.7472},
issn = {0003-6951},
journal = {Applied Physics Letters},
month = {mar},
number = {11},
pages = {111909},
title = {{Anisotropic thermal conductivity in single crystal $\beta$-gallium oxide}},
doi = {10.1063/1.4916078}, url = {http://aip.scitation.org/doi/10.1063/1.4916078},
volume = {106},
year = {2015}
}

@article{Grashchenko2018,
author = {Grashchenko, A. S. and Kukushkin, S. A. and Nikolaev, V. I. and Osipov, A. V. and Osipova, E. V. and Soshnikov, I. P.},
doi = {10.1134/S1063783418050104},
issn = {1063-7834},
journal = {Phys. Solid State},
month = {may},
number = {5},
pages = {852--857},
title = {{Study of the Anisotropic Elastoplastic Properties of $\beta-\mathrm{Ga}_{2}\mathrm{O}_{3}$ Films Synthesized on $\mathrm{SiC/Si}$ Substrates}},
url = {http://link.springer.com/10.1134/S1063783418050104},
volume = {60},
year = {2018}
}

@article{Higashiwaki,
author = {Higashiwaki, Masataka and Sasaki, Kohei and Kuramata, Akito and Masui, Takekazu and Yamakoshi, Shigenobu},
doi = {10.1002/pssa.201330197},
issn = {18626300},
journal = {Phys. Status Solidi A},
month = {jan},
number = {1},
pages = {21--26},
title = {Development of gallium oxide power devices},
url = {http://doi.wiley.com/10.1002/pssa.201330197},
volume = {211},
year = {2014}
}

@article{Higashiwaki2,
author = {Higashiwaki, Masataka and Sasaki, Kohei and Kuramata, Akito and Masui, Takekazu and Yamakoshi, Shigenobu},
doi = {10.1063/1.3674287},
issn = {0003-6951},
journal = {Appl. Phys. Lett.},
month = {jan},
number = {1},
pages = {013504},
title = {Gallium oxide ($\mathrm{Ga}_{2}\mathrm{O}_{3}$) metal-semiconductor field-effect transistors on single-crystal $\beta-\mathrm{Ga}_{2}\mathrm{O}_{3}$ (010) substrates},
url = {http://aip.scitation.org/doi/10.1063/1.3674287},
volume = {100},
year = {2012}
}

@article{Higashiwaki2017,
author = {Higashiwaki, Masataka and Kuramata, Akito and Murakami, Hisashi and Kumagai, Yoshinao},
doi = {10.1088/1361-6463/aa7aff},
issn = {0022-3727},
journal = {J. Phys. D. Appl. Phys.},
month = {aug},
number = {33},
pages = {333002},
publisher = {IOP Publishing},
title = {{State-of-the-art technologies of gallium oxide power devices}},
url = {https://iopscience.iop.org/article/10.1088/1361-6463/aa7aff},
volume = {50},
year = {2017}
}

@article{Handwerg2015,
author = {Handwerg, M. and Mitdank, R. and Galazka, Z. and Fischer, S. F.},
doi = {10.1088/0268-1242/30/2/024006},
issn = {0268-1242},
journal = {Semiconductor Science and Technology},
month = {feb},
number = {2},
pages = {024006},
publisher = {IOP Publishing},
title = {{Temperature-dependent thermal conductivity in Mg-doped and undoped $\beta-\mathrm{Ga}_{2}\mathrm{O}_{3}$ bulk-crystals}},
doi = {10.1088/0268-1242/30/2/024006},
url = {https://iopscience.iop.org/article/10.1088/0268-1242/30/2/024006},
volume = {30},
year = {2015}
}

@article{Handwerg2016,
archivePrefix = {arXiv},
arxivId = {1506.05294},
author = {Handwerg, M. and Mitdank, R. and Galazka, Z. and Fischer, S. F.},
doi = {10.1088/0268-1242/31/12/125006},
eprint = {1506.05294},
issn = {0268-1242},
journal = {Semiconductor Science and Technology},
month = {dec},
number = {12},
pages = {125006},
publisher = {IOP Publishing},
title = {{Temperature-dependent thermal conductivity and diffusivity of a Mg-doped insulating $\beta-\mathrm{Ga}_{2}\mathrm{O}_{3}$ single crystal along [100], [010] and [001]}},
url = {https://iopscience.iop.org/article/10.1088/0268-1242/31/12/125006},
volume = {31},
year = {2016}
}

@article{Irmscher2011,
author = {Irmscher, K. and Galazka, Z. and Pietsch, M. and Uecker, R. and Fornari, R.},
doi = {10.1063/1.3642962},
issn = {0021-8979},
journal = {J. Appl. Phys.},
month = {sep},
number = {6},
pages = {063720},
title = {Electrical properties of $\mathrm{Ga}_{2}\mathrm{O}_{3}$ single crystals grown by the \textrm{Czochralski} method},
url = {http://aip.scitation.org/doi/10.1063/1.3642962},
volume = {110},
year = {2011}
}

@article{Jiang2018,
archivePrefix = {arXiv},
arxivId = {1809.04762},
author = {Jiang, Puqing and Qian, Xin and Li, Xiaobo and Yang, Ronggui},
doi = {10.1063/1.5054573},
eprint = {1809.04762},
issn = {0003-6951},
journal = {Applied Physics Letters},
month = {dec},
number = {23},
pages = {232105},
title = {{Three-dimensional anisotropic thermal conductivity tensor of single crystalline $\ensuremath{\beta}\ensuremath{-}{\mathrm{Ga}}_{2}{\mathrm{O}}_{3}$}},
doi = {10.1063/1.5054573}, url = {http://aip.scitation.org/doi/10.1063/1.5054573},
volume = {113},
year = {2018}
}

@article{Koga1958,
author = {Koga, Issac and Aruga, Masanao and Yoshinaka, Yōichirō},
doi = {10.1103/PhysRev.109.1467},
issn = {0031-899X},
journal = {Phys. Rev.},
month = {mar},
number = {5},
pages = {1467--1473},
title = {{Theory of Plane Elastic Waves in a Piezoelectric Crystalline Medium and Determination of Elastic and Piezoelectric Constants of Quartz}},
url = {https://link.aps.org/doi/10.1103/PhysRev.109.1467},
volume = {109},
year = {1958}
}

@article{Kokubuna2007,
author = {Kokubun, Yoshihiro and Miura, Kasumi and Endo, Fumie and Nakagomi, Shinji},
doi = {10.1063/1.2432946},
issn = {0003-6951},
journal = {Appl. Phys. Lett.},
month = {jan},
number = {3},
pages = {031912},
title = {Sol-gel prepared $\beta-\mathrm{Ga}_{2}\mathrm{O}_{3}$ thin films for ultraviolet photodetectors},
url = {http://aip.scitation.org/doi/10.1063/1.2432946},
volume = {90},
year = {2007}
}

@Article{Kranert2016SciRep,
  author          = {Kranert, Christian and Sturm, Chris and Schmidt-Grund, R{\"{u}}diger and Grundmann, Marius},
  title           = {{Raman tensor elements of $\beta$-Ga2O3}},
  journal         = {Scientific Reports},
  year            = {2016},
  volume          = {6},
  number          = {1},
  pages           = {35964},
  month           = {dec},
  issn            = {2045-2322},
  doi             = {10.1038/srep35964},
  url             = {http://www.nature.com/articles/srep35964},
}

@Article{Kranert2016PRL,
  author   = {Kranert, Christian and Sturm, Chris and Schmidt-Grund, R{\"{u}}diger and Grundmann, Marius},
  title    = {{Raman Tensor Formalism for Optically Anisotropic Crystals}},
  journal  = {Physical Review Letters},
  year     = {2016},
  volume   = {116},
  number   = {12},
  pages    = {127401},
  month    = {mar},
  issn     = {0031-9007},
  doi      = {10.1103/PhysRevLett.116.127401},
  url      = {https://journals.aps.org/prl/pdf/10.1103/PhysRevLett.116.127401},
}

@article{Li2020,
author = {Li, Ruiyang and Liu, Zeyu and Rohskopf, Andrew and Gordiz, Kiarash and Henry, Asegun and Lee, Eungkyu and Luo, Tengfei},
doi = {10.1063/5.0025051},
issn = {0003-6951},
journal = {Applied Physics Letters},
month = {oct},
number = {15},
pages = {152102},
publisher = {AIP Publishing LLC},
title = {{A deep neural network interatomic potential for studying thermal conductivity of $\beta-\mathrm{Ga}_{2}\mathrm{O}_{3}$}},
url = {http://aip.scitation.org/doi/10.1063/5.0025051},
volume = {117},
year = {2020}
}

@article{Luan2019,
author = {Luan, Suzhen and Dong, Linpeng and Jia, Renxu},
doi = {10.1016/j.jcrysgro.2018.09.031},
issn = {00220248},
journal = {J. Cryst. Growth},
number = {September 2018},
pages = {74--81},
publisher = {Elsevier B.V.},
title = {{Analysis of the structural, anisotropic elastic and electronic properties of $\beta-\mathrm{Ga}_{2}\mathrm{O}_{3}$ with various pressures}},
url = {https://doi.org/10.1016/j.jcrysgro.2018.09.031},
volume = {505},
year = {2019}
}

@article{Matsumoto1974,
author = {Matsumoto, Takashi and Aoki, Masaharu and Kinoshita, Akira and Aono, Tomoyoshi},
doi = {10.1143/JJAP.13.1578},
issn = {0021-4922},
journal = {Jpn. J. Appl. Phys.},
month = {oct},
number = {10},
pages = {1578--1582},
publisher = {IOP Publishing},
title = {{Absorption and Reflection of Vapor Grown Single Crystal Platelets of $\beta-\mathrm{Ga}_{2}\mathrm{O}_{3}$}},
url = {https://iopscience.iop.org/article/10.1143/JJAP.13.1578},
volume = {13},
year = {1974}
}

@article{Mock2017,
author = {Mock, Alyssa and Korlacki, Rafa{\l} and Briley, Chad and Darakchieva, Vanya and Monemar, Bo and Kumagai, Yoshinao and Goto, Ken and Higashiwaki, Masataka and Schubert, Mathias},
doi = {10.1103/PhysRevB.96.245205},
issn = {2469-9950},
journal = {Phys. Rev. B},
month = {dec},
number = {24},
pages = {245205},
title = {Band-to-band transitions, selection rules, effective mass, and excitonic contributions in monoclinic \mbox{$\beta$-\mathrm{Ga}_{2}\mathrm{O}_{3}}},
url = {https://link.aps.org/doi/10.1103/PhysRevB.96.245205},
volume = {96},
year = {2017}
}

@article{Mu2017,
author = {Mu, Wenxiang and Jia, Zhitai and Yin, Yanru and Hu, Qiangqiang and Li, Yang and Wu, Baiyi and Zhang, Jian and Tao, Xutang},
doi = {10.1016/j.jallcom.2017.04.185},
issn = {09258388},
journal = {J. Alloys Compd.},
pages = {453--458},
publisher = {Elsevier B.V},
title = {{High quality crystal growth and anisotropic physical characterization of \mbox{$\beta-\mathrm{Ga}_{2}\mathrm{O}_{3}$} single crystals grown by EFG method}},
url = {http://dx.doi.org/10.1016/j.jallcom.2017.04.185},
volume = {714},
year = {2017}
}

@book{Newnham2004,
author = {Newnham, Robert E.},
doi = {10.1093/oso/9780198520757.001.0001},
isbn = {9780198520757},
month = {nov},
publisher = {Oxford University Press},
title = {{Properties of Materials}},
url = {https://academic.oup.com/book/41813},
year = {2004}
}

@article{Nakamura2012,
author = {Nakamura, N. and Ogi, H. and Hirao, M.},
doi = {10.1063/1.3674271},
issn = {0021-8979},
journal = {Journal of Applied Physics},
month = {jan},
number = {1},
pages = {013509},
title = {{Elastic, anelastic, and piezoelectric coefficients of GaN}},
url = {http://aip.scitation.org/doi/10.1063/1.3674271},
volume = {111},
year = {2012}
}

@article{Ogi2002,
author = {Ogi, Hirotsugu and Sato, Keiji and Asada, Takeyasu and Hirao, Masahiko},
doi = {10.1121/1.1512700},
issn = {0001-4966},
journal = {The Journal of the Acoustical Society of America},
month = {dec},
number = {6},
pages = {2553--2557},
pmid = {12508976},
title = {{Complete mode identification for resonance ultrasound spectroscopy}},
url = {http://asa.scitation.org/doi/10.1121/1.1512700},
volume = {112},
year = {2002}
}

@article{Ogi2006,
author = {Ogi, Hirotsugu and Ohmori, Toshinobu and Nakamura, Nobutomo and Hirao, Masahiko},
doi = {10.1063/1.2335684},
issn = {0021-8979},
journal = {Journal of Applied Physics},
month = {sep},
number = {5},
pages = {053511},
title = {{Elastic, anelastic, and piezoelectric coefficients of $\alpha$-quartz determined by resonance ultrasound spectroscopy}},
url = {http://aip.scitation.org/doi/10.1063/1.2335684},
volume = {100},
year = {2006}
}

@article{Onuma2015,
author = {Onuma, Takeyoshi and Saito, Shingo and Sasaki, Kohei and Masui, Tatekazu and Yamaguchi, Tomohiro and Honda, Tohru and Higashiwaki, Masataka},
doi = {10.7567/JJAP.54.112601},
issn = {0021-4922},
journal = {Jpn. J. Appl. Phys.},
month = {nov},
number = {11},
pages = {112601},
title = {{Valence band ordering in $\beta-\mathrm{Ga}_{2}\mathrm{O}_{3}$ studied by polarized transmittance and reflectance spectroscopy}},
url = {https://iopscience.iop.org/article/10.7567/JJAP.54.112601},
volume = {54},
year = {2015}
}

@article{Onuma2016,
author = {Onuma, T. and Saito, S. and Sasaki, K. and Goto, K. and Masui, T. and Yamaguchi, T. and Honda, T. and Kuramata, A. and Higashiwaki, M.},
doi = {10.1063/1.4943175},
issn = {0003-6951},
journal = {Appl. Phys. Lett.},
month = {mar},
number = {10},
pages = {101904},
title = {{Temperature-dependent exciton resonance energies and their correlation with IR-active optical phonon modes in $\beta-\mathrm{Ga}_{2}\mathrm{O}_{3}$ single crystals}},
doi = {10.1063/1.4943175},
url = {http://aip.scitation.org/doi/10.1063/1.4943175},
volume = {108},
year = {2016}
}

@article{Onuma2016b,
author = {Onuma, Takeyoshi and Saito, Shingo and Sasaki, Kohei and Masui, Tatekazu and Yamaguchi, Tomohiro and Honda, Tohru and Kuramata, Akito and Higashiwaki, Masataka},
doi = {10.7567/JJAP.55.1202B2},
issn = {13474065},
journal = {Jpn. J. Appl. Phys.},
number = {12},
title = {{Spectroscopic ellipsometry studies on $\beta-\mathrm{Ga}_{2}\mathrm{O}_{3}$ films and single crystal}},
pages = {1202B2},
volume = {55},
year = {2016}
}

@article{Oshima2008,
author = {Oshima, Takayoshi and Okuno, Takeya and Arai, Naoki and Suzuki, Norihito and Ohira, Shigeo and Fujita, Shizuo},
doi = {10.1143/APEX.1.011202},
issn = {1882-0778},
journal = {Appl. Phys. Express},
month = {jan},
number = {1},
pages = {011202},
title = {Vertical Solar-Blind Deep-Ultraviolet Schottky Photodetectors Based on $\beta-\mathrm{Ga}_{2}\mathrm{O}_{3}$ Substrates},
url = {https://iopscience.iop.org/article/10.1143/APEX.1.011202},
volume = {1},
year = {2008}
}

@article{Pearton,
author = {Pearton, S. J. and Yang, Jiancheng and Cary, Patrick H. and Ren, F. and Kim, Jihyun and Tadjer, Marko J. and Mastro, Michael A.},
doi = {10.1063/1.5006941},
issn = {1931-9401},
journal = {Appl. Phys. Rev.},
month = {mar},
number = {1},
pages = {011301},
title = {A review of $\mathrm{Ga}_{2}\mathrm{O}_{3}$ materials, processing, and devices},
doi = {10.1063/1.5006941},
url = {http://aip.scitation.org/doi/10.1063/1.5006941},
volume = {5},
year = {2018}
}

@article{Pearton2018,
author = {Pearton, S. J. and Ren, Fan and Tadjer, Marko and Kim, Jihyun},
doi = {10.1063/1.5062841},
issn = {0021-8979},
journal = {J. Appl. Phys.},
month = {dec},
number = {22},
pages = {220901},
title = {{Perspective: $\mathrm{Ga}_{2}\mathrm{O}_{3}$ for ultra-high power rectifiers and \textrm{MOSFETS}}},
doi = {10.1063/1.5062841},
url = {http://aip.scitation.org/doi/10.1063/1.5062841},
volume = {124},
year = {2018}
}

@article{Peelaers2015,
author = {Peelaers, Hartwin and Van de Walle, Chris G.},
doi = {10.1002/pssb.201451551},
issn = {03701972},
journal = {Phys. Status Solidi B},
month = {apr},
number = {4},
pages = {828--832},
title = {{Brillouin zone and band structure of $\beta-\mathrm{Ga}_{2}\mathrm{O}_{3}$ }},
url = {https://onlinelibrary.wiley.com/doi/10.1002/pssb.201451551},
volume = {252},
year = {2015}
}

@article{Perez2022,
author = {P{\'{e}}rez, Luis A. and Xu, Kai and Wagner, Markus R. and D{\"{o}}rling, Bernhard and Perevedentsev, Aleksandr and Go{\~{n}}i, Alejandro R. and Campoy-Quiles, Mariano and Alonso, M. Isabel and Reparaz, Juan Sebasti{\'{a}}n},
doi = {10.1063/5.0066166},
issn = {0034-6748},
journal = {Review of Scientific Instruments},
month = {mar},
number = {3},
pages = {034902},
pmid = {35365009},
publisher = {AIP Publishing LLCAIP Publishing},
title = {{Anisotropic thermoreflectance thermometry: A contactless frequency-domain thermoreflectance approach to study anisotropic thermal transport}},
url = {https://aip.scitation.org/doi/abs/10.1063/5.0066166},
volume = {93},
year = {2022}
}

@article{Ratnaparkhe2017,
author = {Ratnaparkhe, Amol and Lambrecht, Walter R. L.},
doi = {10.1063/1.4978668},
issn = {0003-6951},
journal = {Appl. Phys. Lett.},
month = {mar},
number = {13},
pages = {132103},
title = {{Quasiparticle self-consistent GW band structure of $\mathrm{Ga}_{2}\mathrm{O}_{3}$ and the anisotropy of the absorption onset}},
url = {http://aip.scitation.org/doi/10.1063/1.4978668},
volume = {110},
year = {2017}
}

@article{Ricci2016,
author = {Ricci, F. and Boschi, F. and Baraldi, A. and Filippetti, A. and Higashiwaki, M. and Kuramata, A. and Fiorentini, V. and Fornari, R.},
doi = {10.1088/0953-8984/28/22/224005},
issn = {1361648X},
journal = {J. Phys. Condens. Matter},
number = {22},
pages = {224005},
publisher = {IOP Publishing},
title = {{Theoretical and experimental investigation of optical absorption anisotropy in $\mathrm{Ga}_{2}\mathrm{O}_{3}$}},
url = {http://dx.doi.org/10.1088/0953-8984/28/22/224005},
volume = {28},
year = {2016}
}

@article{Santia2015,
author = {Santia, Marco D. and Tandon, Nandan and Albrecht, J. D.},
doi = {10.1063/1.4927742},
issn = {0003-6951},
journal = {Applied Physics Letters},
month = {jul},
number = {4},
pages = {041907},
title = {{Lattice thermal conductivity in $\beta-\mathrm{Ga}_{2}\mathrm{O}_{3}$ from first principles}},
doi = {10.1063/1.4927742},
url = {http://aip.scitation.org/doi/10.1063/1.4927742},
volume = {107},
year = {2015}
}

@article{Slomski2017,
author = {Slomski, M. and Blumenschein, N. and Paskov, P. P. and Muth, J. F. and Paskova, T.},
doi = {10.1063/1.4986478},
issn = {0021-8979},
journal = {Journal of Applied Physics},
month = {jun},
number = {23},
pages = {235104},
title = {{Anisotropic thermal conductivity of $\beta-\mathrm{Ga}_{2}\mathrm{O}_{3}$ at elevated temperatures: Effect of Sn and Fe dopants}},
doi = {10.1063/1.4986478},
url = {http://aip.scitation.org/doi/10.1063/1.4986478},
volume = {121},
year = {2017}
}

@article{Smith1971,
author = {Smith, R. T. and Welsh, F. S.},
doi = {10.1063/1.1660528},
issn = {00218979},
journal = {J. Appl. Phys.},
number = {6},
pages = {2219--2230},
title = {{Temperature dependence of the elastic, piezoelectric, and dielectric constants of lithium tantalate and lithium niobate}},
volume = {42},
year = {1971}
}

@article{Spencer2022,
author = {Spencer, Joseph A. and Mock, Alyssa L. and Jacobs, Alan G. and Schubert, Mathias and Zhang, Yuhao and Tadjer, Marko J.},
doi = {10.1063/5.0078037},
issn = {19319401},
journal = {Applied Physics Reviews},
number = {1},
publisher = {AIP Publishing LLC},
title = {{A review of band structure and material properties of transparent conducting and semiconducting oxides: $\mathrm{Ga}_{2}\mathrm{O}_{3}$, $\mathrm{Al}_{2}\mathrm{O}_{3}$, $\mathrm{In}_{2}\mathrm{O}_{3}$, $\mathrm{ZnO}$, $\mathrm{Sn}\mathrm{O}_{2}$, $\mathrm{CdO}$, $\mathrm{NiO}$, $\mathrm{CuO}$, and $\mathrm{Sc}_{2}\mathrm{O}_{3}$}},
volume = {9},
year = {2022}
}

@article{Sturm2015,
    author = {Sturm, C. and Furthmüller, J. and Bechstedt, F. and Schmidt-Grund, R. and Grundmann, M.},
    title = {Dielectric tensor of monoclinic Ga2O3 single crystals in the spectral range 0.5–8.5 eV},
    journal = {APL Materials},
    volume = {3},
    number = {10},
    pages = {106106},
    year = {2015},
    month = {10},
    issn = {2166-532X},
    doi = {10.1063/1.4934705},
}

@article{Sturm2016b,
archivePrefix = {arXiv},
arxivId = {1601.03760},
author = {Sturm, Chris and Grundmann, Marius},
doi = {10.1103/PhysRevA.93.053839},
eprint = {1601.03760},
issn = {2469-9926},
journal = {Phys. Rev. A},
month = {may},
number = {5},
pages = {053839},
title = {{Singular optical axes in biaxial crystals and analysis of their spectral dispersion effects in \textbf{\mbox{$\mathrm{Ga}_{2}\mathrm{O}_{3}$}} }},
url = {https://link.aps.org/doi/10.1103/PhysRevA.93.053839},
volume = {93},
year = {2016}
}

@article{Ueda1997,
author = {Ueda, Naoyuki and Hosono, Hideo and Waseda, Ryuta and Kawazoe, Hiroshi},
doi = {10.1063/1.119693},
issn = {0003-6951},
journal = {Appl. Phys. Lett.},
month = {aug},
number = {7},
pages = {933--935},
title = {Anisotropy of electrical and optical properties in $\beta-\mathrm{Ga}_{2}\mathrm{O}_{3}$ single crystals},
url = {http://aip.scitation.org/doi/10.1063/1.119693},
volume = {71},
year = {1997}
}

@article{Varley2015,
author = {Varley, Joel B. and Schleife, Andr{\'{e}}},
doi = {10.1088/0268-1242/30/2/024010},
issn = {0268-1242},
journal = {Semicond. Sci. Technol.},
month = {feb},
number = {2},
pages = {024010},
publisher = {IOP Publishing},
title = {{Bethe–Salpeter calculation of optical-absorption spectra of $\mathrm{In}_{2}\mathrm{O}_{3}$ and $\mathrm{Ga}_{2}\mathrm{O}_{3}$}},
url = {https://iopscience.iop.org/article/10.1088/0268-1242/30/2/024010},
volume = {30},
year = {2015}
}

@article{Yan2018,
author = {Yan, Zhequan and Kumar, Satish},
doi = {10.1039/C8CP05139A},
issn = {1463-9076},
journal = {Physical Chemistry Chemical Physics},
number = {46},
pages = {29236--29242},
pmid = {30427340},
publisher = {Royal Society of Chemistry},
title = {{Phonon mode contributions to thermal conductivity of pristine and defective $\beta-\mathrm{Ga}_{2}\mathrm{O}_{3}$}},
url = {http://xlink.rsc.org/?DOI=C8CP05139A},
volume = {20},
year = {2018}
}

@article{Yamaguchi2004,
author = {Yamaguchi, Kenji},
doi = {10.1016/j.ssc.2004.07.030},
issn = {00381098},
journal = {Solid State Commun.},
month = {sep},
number = {12},
pages = {739--744},
publisher = {Elsevier Ltd},
title = {{First principles study on electronic structure of $\beta$-Ga 2O3}},
volume = {131},
year = {2004}
}

@article{Zhang2020,
author = {Zhang, Yingying and Su, Qun and Zhu, Jie and Koirala, Sandhaya and Koester, Steven J. and Wang, Xiaojia},
doi = {10.1063/5.0004984},
issn = {0003-6951},
journal = {Applied Physics Letters},
month = {may},
number = {20},
pages = {202101},
publisher = {AIP Publishing LLC},
title = {{Thickness-dependent thermal conductivity of mechanically exfoliated $\beta-\mathrm{Ga}_{2}\mathrm{O}_{3}$ thin films}},
url = {http://aip.scitation.org/doi/10.1063/5.0004984},
volume = {116},
year = {2020}
}

@article{Zhang2021optical,
author = {Zhang, Naiji and Kislyakov, Ivan M. and Xia, Changtai and Qi, Hongji and Wang, Jun and Mohamed, H. F.},
doi = {10.1364/OE.427021},
issn = {1094-4087},
journal = {Opt. Express},
month = {jun},
number = {12},
pages = {18587},
pmid = {34154112},
title = {{Anisotropic luminescence and third-order electric susceptibility of $\mathrm{Mg}$-doped gallium oxide under the half-bandgap edge}},
url = {https://doi.org/10.1364/OE.427021},
volume = {29},
year = {2021}
}

@article{speziale2014brillouin,
  title={Brillouin scattering and its application in geosciences},
  author={Speziale, Sergio and Marquardt, Hauke and Duffy, Thomas S},
  journal={Reviews in Mineralogy and Geochemistry},
  volume={78},
  number={1},
  pages={543--603},
  year={2014},
  publisher={GeoScienceWorld}
}

@book{nye1985physical,
  title={Physical properties of crystals: their representation by tensors and matrices},
  author={Nye, John Frederick and others},
  year={1985},
  publisher={Oxford university press}
}

@article{graczykowski2019brillouin,
  title={Brillouin light scattering under one-dimensional confinement: Symmetry and interference self-canceling},
  author={Graczykowski, Bartlomiej and Gueddida, A and Djafari-Rouhani, Bahram and Butt, H-J and Fytas, Georg},
  journal={Physical Review B},
  volume={99},
  number={16},
  pages={165431},
  year={2019},
  publisher={APS}
}

@article{vacher1972brillouin,
  title={Brillouin scattering: A tool for the measurement of elastic and photoelastic constants},
  author={Vacher, R and Boyer, L},
  journal={Physical Review B},
  volume={6},
  number={2},
  pages={639},
  year={1972},
  publisher={APS}
}

@article{babacic2021thickness,
  title={Thickness-Dependent Elastic Softening of Few-Layer Free-Standing MoSe2},
  author={Babacic, Visnja and Saleta Reig, David and Varghese, Sebin and Vasileiadis, Thomas and Coy, Emerson and Tielrooij, Klaas-Jan and Graczykowski, Bartlomiej},
  journal={Advanced Materials},
  volume={33},
  number={23},
  pages={2008614},
  year={2021},
  publisher={Wiley Online Library}
}

@article{grundmann2020principal,
  title={The principal axes systems for the elastic properties of monoclinic gallia},
  author={Grundmann, Marius},
  journal={Scientific Reports},
  volume={10},
  number={1},
  pages={19486},
  year={2020},
  publisher={Nature Publishing Group},
  url={https://www.nature.com/articles/s41598-020-73970-9}
}

@article{adachi2018unusual,
  title={Unusual elasticity of monoclinic $\beta-\mathrm{Ga}_{2}\mathrm{O}_{3}$},
  author={Adachi, K and Ogi, H and Takeuchi, N and Nakamura, N and Watanabe, H and Ito, T and Ozaki, Y},
  journal={Journal of Applied Physics},
  volume={124},
  number={8},
  pages={085102},
  year={2018},
  publisher={AIP Publishing LLC}
}

@article{miller2017numerical,
  title={Numerical modelling of the Czochralski growth of $\beta$-Ga2O3},
  author={Miller, Wolfram and B{\"o}ttcher, Klaus and Galazka, Zbigniew and Schreuer, J{\"u}rgen},
  journal={Crystals},
  volume={7},
  number={1},
  pages={26},
  year={2017},
  publisher={MDPI}
}

@article{doi:10.1063/1.4812323,
author = {Jain,Anubhav  and Ong,Shyue Ping  and Hautier,Geoffroy  and Chen,Wei  and Richards,William Davidson  and Dacek,Stephen  and Cholia,Shreyas  and Gunter,Dan  and Skinner,David  and Ceder,Gerbrand  and Persson,Kristin A. },
title = {Commentary: The Materials Project: A materials genome approach to accelerating materials innovation},
journal = {APL Materials},
volume = {1},
number = {1},
pages = {011002},
year = {2013},
doi = {10.1063/1.4812323},

URL = { 
        https://doi.org/10.1063/1.4812323
    
},
eprint = { 
        https://doi.org/10.1063/1.4812323
    
}

}

@article{mouhat2014necessary,
  title={Necessary and sufficient elastic stability conditions in various crystal systems},
  author={Mouhat, F{\'e}lix and Coudert, Fran{\c{c}}ois-Xavier},
  journal={Physical review B},
  volume={90},
  number={22},
  pages={224104},
  year={2014},
  publisher={APS}
}

@article{CarreteCPC17,
  title      = {almaBTE: A solver of the space-time dependent Boltzmann
                transport equation for phonons in structured materials},
  journal    = {Comput. Phys. Commun.},
  volume     = {220},
  pages      = {351--362},
  year       = {2017},
  author     = {Jes\'us Carrete and Bjorn Vermeersch and Ankita Katre
                and Ambroise van Roekeghem and Tao Wang and
                Georg K.H. Madsen and Natalio Mingo},
}

@Article{PerdewPRL96,
  title     = {Generalized Gradient Approximation Made Simple},
  author    = {Perdew, John P. and Burke, Kieron and Ernzerhof, Matthias},
  journal   = {Phys. Rev. Lett.},
  volume    = {77},
  number    = {18},
  pages     = {3865--3868},
  year      = {1996},
}

@Article{KressePRB99,
  author    = {G. Kresse and D. Joubert},
  title     = {From ultrasoft pseudopotentials to the projector
               augmented-wave method},
  journal   = {Phys. Rev. B},
  year      = {1999},
  volume    = {59},
  pages     = {1758},
}

@article{KressePRB96,
  title      = {Efficient iterative schemes for ab initio total-energy calculations using a plane-wave basis set},
  author     = {Kresse, Georg and Furthm{\"u}ller, J{\"u}rgen},
  journal    = {Phys. Rev. B},
  volume     = {54},
  number     = {16},
  pages      = {11169},
  year       = {1996},
}

@article{TogoSM15,
  journal    = {Scr. Mater.},
  year       = {2015},
  title      = {First principles phonon calculations in materials science},
  author     = {Togo, A and Tanaka, I},
  pages      = {1--5},
  volume     = {108},
}

@article{LiCPC14,
  author     = {Wu Li and Jes\'us Carrete and Nebil A. Katcho and Natalio Mingo},
  title      = {{ShengBTE:} a solver of the {B}oltzmann transport equation for phonons},
  journal    = {Comp. Phys. Commun.},
  volume     = {185},
  pages      = {1747--1758},
  year       = {2014}
}

@article{aiGK,
  year = {2017},
  title = {{Ab Initio Green-Kubo Approach for the Thermal Conductivity of Solids
           }},
  author = {Carbogno, Christian and Ramprasad, Rampi and Scheffler, Matthias},
  journal = {Physical Review Letters},
  issn = {0031-9007},
  doi = {10.1103/physrevlett.118.175901},
  pmid = {28498695},
  eprint = {1608.06917},
  pages = {175901},
  number = {17},
  volume = {118},
}

@article{Knoop2023_PRB,
  year = {2023},
  title = {{Ab initio Green-Kubo simulations of heat transport in solids: Method
           and implementation}},
  author = {Knoop, Florian and Scheffler, Matthias and Carbogno, Christian},
  journal = {Physical Review B},
  issn = {2469-9950},
  doi = {10.1103/physrevb.107.224304},
  eprint = {2209.01139},
  pages = {224304},
  number = {22},
  volume = {107},
}

@article{Knoop2023_PRL,
  year = {2023},
  title = {{Anharmonicity in Thermal Insulators: An Analysis from First
           Principles}},
  author = {Knoop, Florian and Purcell, Thomas A. R. and Scheffler, Matthias and
            Carbogno, Christian},
  journal = {Physical Review Letters},
  issn = {0031-9007},
  doi = {10.1103/physrevlett.130.236301},
  pmid = {37354415},
  pages = {236301},
  number = {23},
  volume = {130},
}

@article{fhi-aims,
  year = {2009},
  title = {{Ab initio molecular simulations with numeric atom-centered orbitals}
           },
  author = {Blum, Volker and Gehrke, Ralf and Hanke, Felix and Havu, Paula and
            Havu, Ville and Ren, Xinguo and Reuter, Karsten and Scheffler,
            Matthias},
  journal = {Computer Physics Communications},
  issn = {0010-4655},
  doi = {10.1016/j.cpc.2009.06.022},
  pages = {2175--2196},
  number = {11},
  volume = {180},
}

@article{fhi-vibes,
  year = {2020},
  title = {{FHI-vibes: Ab Initio Vibrational Simulations}},
  author = {Knoop, Florian and Purcell, Thomas and Scheffler, Matthias and
            Carbogno, Christian},
  journal = {Journal of Open Source Software},
  doi = {10.21105/joss.02671},
  pages = {2671},
  number = {56},
  volume = {5},
}

@article{TDEP2nd,
  year = {2013},
  title = {{Temperature dependent effective potential method for accurate free
           energy calculations of solids}},
  author = {Hellman, Olle and Steneteg, Peter and Abrikosov, I. A. and Simak, S.
            I.},
  journal = {Physical Review B},
  issn = {1098-0121},
  doi = {10.1103/physrevb.87.104111},
  eprint = {1303.1145},
  pages = {104111},
  number = {10},
  volume = {87},
}

@article{TDEP2024,
  year = {2024},
  title = {{TDEP: Temperature Dependent Effective Potentials}},
  author = {Knoop, Florian and Shulumba, Nina and Castellano, Aloïs and Batista,
            J P Alvarinhas and Farris, Roberta and Verstraete, Matthieu J and
            Heine, Matthew and Broido, David and Kim, Dennis S and Klarbring,
            Johan and Abrikosov, Igor A and Simak, Sergei I and Hellman, Olle},
  journal = {Journal of Open Source Software},
  doi = {10.21105/joss.06150},
  pages = {6150},
  number = {94},
  volume = {9},
}

@article{stresses,
  year = {2015},
  title = {{All-electron formalism for total energy strain derivatives and
           stress tensor components for numeric atom-centered orbitals}},
  author = {Knuth, Franz and Carbogno, Christian and Atalla, Viktor and Blum,
            Volker and Scheffler, Matthias},
  journal = {Computer Physics Communications},
  issn = {0010-4655},
  doi = {10.1016/j.cpc.2015.01.003},
  pages = {33--50},
  volume = {190},
}

@article{Zhao2025,
  year = {2026},
  title = {{Elucidating the Role of Polymorphism for Anisotropic Thermal Conductivities in Ga$_2$O$_3$ Using Fully Anharmonic Green-Kubo Calculations}},
  author = {Zhao, Shuo and Purcell, Thomas A. R. and Kang, Kisung and Scheffler, Matthias and
            Carbogno, Christian},
  journal = {in preparation},
}

@article{McGaughey.2004, 
year = {2004}, 
title = {{Quantitative validation of the Boltzmann transport equation phonon thermal conductivity model under the single-mode relaxation time approximation}}, 
author = {McGaughey, AJH and Kaviany, M}, 
journal = {Physical Review-Section B-Condensed Matter}, 
doi = {10.1103/physrevb.69.094303}, 
pages = {094303}, 
number = {9}, 
volume = {69}, 
language = {English}
}

@article{Simoncelli.2022, 
year = {2022}, 
title = {{Wigner Formulation of Thermal Transport in Solids}}, 
author = {Simoncelli, Michele and Marzari, Nicola and Mauri, Francesco}, 
journal = {Physical Review X}, 
doi = {10.1103/physrevx.12.041011}, 
pages = {041011}, 
number = {4}, 
volume = {12}
}

@article{Carbogno.2022, 
year = {2022}, 
title = {{Numerical quality control for DFT-based materials databases}}, 
author = {Carbogno, Christian and Thygesen, Kristian Sommer and Bieniek, Björn and Draxl, Claudia and Ghiringhelli, Luca M. and Gulans, Andris and Hofmann, Oliver T. and Jacobsen, Karsten W. and Lubeck, Sven and Mortensen, Jens Jørgen and Strange, Mikkel and Wruss, Elisabeth and Scheffler, Matthias}, 
journal = {npj Computational Materials}, 
doi = {10.1038/s41524-022-00744-4}, 
pages = {69}, 
number = {1}, 
volume = {8}
}

@article{Esfarjani.2008, 
year = {2008}, 
title = {{Method to extract anharmonic force constants from first principles calculations}}, 
author = {Esfarjani, K and Stokes, H T}, 
journal = {Physical Review B}, 
doi = {10.1103/physrevb.77.144112}, 
pages = {144112}, 
volume = {77}
}

@article{fainozzi_stimulated_2024,
	year = {2024},
    title = {Stimulated Brillouin Scattering in the Time Domain at 1 nm − 1 Wave Vector},
	volume = {132},
	issn = {0031-9007, 1079-7114},
	url = {https://link.aps.org/doi/10.1103/PhysRevLett.132.033802},
	doi = {10.1103/PhysRevLett.132.033802},
	pages = {033802},
	number = {3},
	journal = {Physical Review Letters},
	author = {Fainozzi, Danny and Foglia, Laura and Khatu, Nupur N. and Masciovecchio, Claudio and Mincigrucci, Riccardo and Paltanin, Ettore and Bencivenga, Filippo},
	urldate = {2024-03-14},
	date = {2024-01-17},
	langid = {english},
}

@article{foglia_extreme_2023,
	year = {2023},
    title = {Extreme ultraviolet transient gratings: A tool for nanoscale photoacoustics},
	volume = {29},
	issn = {22135979},
	url = {https://linkinghub.elsevier.com/retrieve/pii/S221359792300006X},
	doi = {10.1016/j.pacs.2023.100453},
	shorttitle = {Extreme ultraviolet transient gratings},
	pages = {100453},
	journal  = {Photoacoustics},
	author = {Foglia, L. and Mincigrucci, R. and Maznev, A.A. and Baldi, G. and Capotondi, F. and Caporaletti, F. and Comin, R. and De Angelis, D. and Duncan, R.A. and Fainozzi, D. and Kurdi, G. and Li, J. and Martinelli, A. and Masciovecchio, C. and Monaco, G. and Milloch, A. and Nelson, K.A. and Occhialini, C.A. and Pancaldi, M. and Pedersoli, E. and Pelli-Cresi, J.S. and Simoncig, A. and Travasso, F. and Wehinger, B. and Zanatta, M. and Bencivenga, F.},
	urldate = {2024-03-14},
	date = {2023-02},
	langid = {english},
}

@article{Chen2023, 
  year     = {2023}, 
  title    = {Anomalous Temperature-Dependent Phonon Anharmonicity and Strain Engineering of Thermal Conductivity in β‑Ga2O3}, 
  author   = {Chen, Ying and Peng, Lei and Wu, Yu and Ma, Congcong and Wu, Ao and Zhang, Hao and Fang, Zhilai}, 
  journal  = {The Journal of Physical Chemistry C}, 
  issn     = {1932-7447}, 
  doi      = {10.1021/acs.jpcc.3c02866}, 
  pages    = {13356--13363}, 
  number   = {27}, 
  volume   = {127}
}

@article{Yang2024, 
  year     = {2024}, 
  title    = {Lattice thermal conductivity of β-, α- and κ- Ga2O3: a first-principles computational study}, 
  author   = {Yang, Jinfeng and Xu, Yongze and Wang, Xiaonan and Zhang, Xu and He, Yang and Sun, Huarui}, 
  journal  = {Applied Physics Express}, 
  issn     = {1882-0778}, 
  doi      = {10.35848/1882-0786/ad0ba8}, 
  pages    = {011001}, 
  number   = {1}, 
  volume   = {17}
}

@article{Sun2023, 
  year     = {2023}, 
  title    = {A neuroevolution potential for predicting the thermal conductivity of α, β, and ε-Ga2O3}, 
  author   = {Sun, Zhanpeng and Qi, Zijun and Liang, Kang and Sun, Xiang and Zhang, Zhaofu and Li, Lijie and Wang, Qijun and Zhang, Guoqing and Wu, Gai and Shen, Wei}, 
  journal  = {Applied Physics Letters}, 
  issn     = {0003-6951}, 
  doi      = {10.1063/5.0165320}, 
  pages    = {192202}, 
  number   = {19}, 
  volume   = {123}
}

@article{xu_-plane_2023,
    title = {In-plane thermal diffusivity determination using beam-offset frequency-domain thermoreflectance with a one-dimensional optical heat source},
    volume = {214},
    issn = {0017-9310},
    url = {https://www.sciencedirect.com/science/article/pii/S0017931023005227},
    doi = {10.1016/j.ijheatmasstransfer.2023.124376},
    urldate = {2025-09-26},
    journal = {International Journal of Heat and Mass Transfer},
    author = {Xu, Kai and Guo, Jiali and Raciti, Grazia and Goni, Alejandro R. and Alonso, M. Isabel and Borrisé, Xavier and Zardo, Ilaria and Campoy-Quiles, Mariano and Reparaz, Juan Sebastián},
    month = nov,
    year = {2023},
    pages = {124376},
}

@article{bencivenga_nanoscale_2019,
    title = {Nanoscale transient gratings excited and probed by extreme ultraviolet femtosecond pulses},
    volume = {5},
    url = {https://www.science.org/doi/full/10.1126/sciadv.aaw5805},
    doi = {10.1126/sciadv.aaw5805},
    number = {7},
    urldate = {2024-09-19},
    journal = {Science Advances},
    author = {Bencivenga, F. and Mincigrucci, R. and Capotondi, F. and Foglia, L. and Naumenko, D. and Maznev, A. A. and Pedersoli, E. and Simoncig, A. and Caporaletti, F. and Chiloyan, V. and Cucini, R. and Dallari, F. and Duncan, R. A. and Frazer, T. D. and Gaio, G. and Gessini, A. and Giannessi, L. and Huberman, S. and Kapteyn, H. and Knobloch, J. and Kurdi, G. and Mahne, N. and Manfredda, M. and Martinelli, A. and Murnane, M. and Principi, E. and Raimondi, L. and Spampinati, S. and Spezzani, C. and Trovò, M. and Zangrando, M. and Chen, G. and Monaco, G. and Nelson, K. A. and Masciovecchio, C.},
    month = jul,
    year = {2019},
    note = {Publisher: American Association for the Advancement of Science},
    pages = {eaaw5805},
}

@article{bencivenga_extreme_2023,
    title = {Extreme ultraviolet transient gratings},
    volume = {8},
    issn = {null},
    url = {https://doi.org/10.1080/23746149.2023.2220363},
    doi = {10.1080/23746149.2023.2220363},
    number = {1},
    urldate = {2024-09-19},
    journal = {Advances in Physics: X},
    author = {Bencivenga, F. and Capotondi, F. and Foglia, L. and Mincigrucci, R. and Masciovecchio, C.},
    month = dec,
    year = {2023},
    note = {Publisher: Taylor \& Francis
\_eprint: https://doi.org/10.1080/23746149.2023.2220363},
    pages = {2220363},
}

@article{Klimm2022,
author = {Klimm, Detlef and Amgalan, Binderiya and Ganschow, Steffen and Kwasniewski, Albert and Galazka, Zbigniew and Bickermann, Matthias},
title = {The Thermal Conductivity Tensor of β-Ga2O3 from 300 to 1275 K},
journal = {Crystal Research and Technology},
volume = {58},
number = {2},
pages = {2200204},
doi = {https://doi.org/10.1002/crat.202200204},
url = {https://onlinelibrary.wiley.com/doi/abs/10.1002/crat.202200204},
eprint = {https://onlinelibrary.wiley.com/doi/pdf/10.1002/crat.202200204},
year = {2023}
}

@article{Ahrling2024,
  title = {Resistive and ballistic phonon transport in \ensuremath{\beta}-${\mathrm{Ga}}_{2}{\mathrm{O}}_{3}$},
  author = {Ahrling, R. and Mitdank, R. and Popp, A. and Rehm, J. and Akhtar, A. and Galazka, Z. and Fischer, S. F.},
  journal = {Phys. Rev. B},
  volume = {110},
  issue = {8},
  pages = {085302},
  numpages = {9},
  year = {2024},
  month = {Aug},
  publisher = {American Physical Society},
  doi = {10.1103/PhysRevB.110.085302},
  url = {https://link.aps.org/doi/10.1103/PhysRevB.110.085302}
}

@article{Rybin2024,
    author = {Rybin, Nikita and Shapeev, Alexander},
    title = {A moment tensor potential for lattice thermal conductivity calculations of α and β phases of Ga2O3},
    journal = {Journal of Applied Physics},
    volume = {135},
    number = {20},
    pages = {205108},
    year = {2024},
    month = {05},
    issn = {0021-8979},
    doi = {10.1063/5.0206215},
    url = {https://doi.org/10.1063/5.0206215},
}

@article{Tu2025,
title = {Unveiling the impact of four-phonon scattering on thermal transport properties of the bulk β-Ga2O3 and monolayer Ga2O3},
journal = {Physica E: Low-dimensional Systems and Nanostructures},
volume = {165},
pages = {116099},
year = {2025},
issn = {1386-9477},
doi = {https://doi.org/10.1016/j.physe.2024.116099},
url = {https://www.sciencedirect.com/science/article/pii/S1386947724002030},
author = {Hui Tu and Yuxiong Xue and Rongxing Cao and Yang Liu and Shu Zheng and Hongxia Li and Yuting Guo and Haiyi Sun and Dan Han}
}

@article{Safieddine2024,
title = {Theoretical investigation of the thermal conductivity of Ga2O3 polymorphs},
journal = {Solid State Communications},
volume = {394},
pages = {115715},
year = {2024},
issn = {0038-1098},
doi = {https://doi.org/10.1016/j.ssc.2024.115715},
url = {https://www.sciencedirect.com/science/article/pii/S0038109824002928},
author = {Fatima Safieddine and Fouad {El Haj Hassan} and Michel Kazan}
}

@article{meissner:2024anisotropy_PLE_bGaO,
  title={Anisotropy of optical transitions in $\beta$-Ga2O3 investigated by polarized photoluminescence excitation spectroscopy},
  author={Mei{\ss}ner, Moritz and Bernhardt, Nils and Nippert, Felix and Janzen, Benjamin M and Galazka, Zbigniew and Wagner, Markus R},
  journal={Applied Physics Letters},
  volume={124},
  number={15},
  year={2024},
  publisher={AIP Publishing}
}

@article{Ahman1996,
author = "{\AA}hman, J. and Svensson, G. and Albertsson, J.",
title = "{A Reinvestigation of {$\beta$}-Gallium Oxide }",
journal = "Acta Crystallographica Section C",
year = "1996",
volume = "52",
number = "6",
pages = "1336--1338",
month = "Jun",
doi = {10.1107/S0108270195016404},
url = {https://doi.org/10.1107/S0108270195016404},
}

@article{King1958,
author = {King, E. G.},
title = {Low Temperature Heat Capacities and Entropies at 298.15°K. of Some Oxides of Gallium, Germanium, Molybdenum and Niobium},
journal = {Journal of the American Chemical Society},
volume = {80},
number = {8},
pages = {1799-1800},
year = {1958},
doi = {10.1021/ja01541a005},
URL = {https://doi.org/10.1021/ja01541a005
},
eprint = {https://doi.org/10.1021/ja01541a005
}
}

@article{Minnich2012PRL,
  author  = {Minnich, A. J.},
  title   = {Determining Phonon Mean Free Paths from Observations of Quasiballistic Thermal Transport},
  journal = {Physical Review Letters},
  volume  = {109},
  number  = {20},
  pages   = {205901},
  year    = {2012},
  doi     = {10.1103/PhysRevLett.109.205901}
}

@article{Hua2015JAP,
  author  = {Hua, Chengyun and Minnich, Austin J.},
  title   = {Semi-analytical solution to the frequency-dependent Boltzmann transport equation for cross-plane heat conduction in thin films},
  journal = {Journal of Applied Physics},
  volume  = {117},
  number  = {17},
  pages   = {175306},
  year    = {2015},
  doi     = {10.1063/1.4919432}
}

@article{Cuffe2015PRB,
  author  = {Cuffe, John and Eliason, Jeffrey K. and Maznev, A. A. and Collins, K. C. and Johnson, J. A. and Shchepetov, Andrey and Prunnila, Mika and Ahopelto, Jouni and Sotomayor Torres, Clivia M. and Chen, Gang and Nelson, Keith A.},
  title   = {Reconstructing phonon mean-free-path contributions to thermal conductivity using nanoscale membranes},
  journal = {Physical Review B},
  volume  = {91},
  number  = {24},
  pages   = {245423},
  year    = {2015},
  doi     = {10.1103/PhysRevB.91.245423}
}

@article{Zeng2019JAP,
  author  = {Zeng, Yuqiang and Marconnet, Amy},
  title   = {Reevaluating the suppression function for phonon transport in nanostructures by Monte Carlo techniques},
  journal = {Journal of Applied Physics},
  volume  = {125},
  number  = {3},
  pages   = {034301},
  year    = {2019},
  doi     = {10.1063/1.5048626}
}

@article{Mincigrucci2018NIMA,
  author  = {Mincigrucci, R. and Foglia, L. and Naumenko, D. and Pedersoli, E. and Simoncig, A. and Cucini, R. and Gessini, A. and Kiskinova, M. and Kurdi, G. and Mahne, N. and Manfredda, M. and Nikolov, I. P. and Principi, E. and Raimondi, L. and Zangrando, M. and Masciovecchio, C. and Capotondi, F. and Bencivenga, F.},
  title   = {Advances in instrumentation for FEL-based four-wave-mixing experiments},
  journal = {Nuclear Instruments and Methods in Physics Research, Section A: Accelerators, Spectrometers, Detectors and Associated Equipment},
  volume  = {907},
  pages   = {132--148},
  year    = {2018},
  doi     = {10.1016/j.nima.2018.03.051}
}

@article{tamura_isotope_1983,
    title = {Isotope scattering of dispersive phonons in {Ge}},
    volume = {27},
    url = {https://link.aps.org/doi/10.1103/PhysRevB.27.858},
    doi = {10.1103/PhysRevB.27.858},
    number = {2},
    urldate = {2026-07-10},
    journal = {Physical Review B},
    publisher = {American Physical Society},
    author = {Tamura, Shin-ichiro},
    month = jan,
    year = {1983},
    pages = {858--866},
}

@article{kargar_advances_2021,
    title = {Advances in {Brillouin}–{Mandelstam} light-scattering spectroscopy},
    volume = {15},
    copyright = {2021 Springer Nature Limited},
    issn = {1749-4893},
    url = {https://www.nature.com/articles/s41566-021-00836-5},
    doi = {10.1038/s41566-021-00836-5},
    language = {en},
    number = {10},
    urldate = {2026-07-17},
    journal = {Nature Photonics},
    publisher = {Nature Publishing Group},
    author = {Kargar, Fariborz and Balandin, Alexander A.},
    month = oct,
    year = {2021},
    pages = {720--731},
}

@article{wright_acoustic_2025,
    title = {Acoustic phonon characteristics of (001) and ( 2¯01)  β-{Ga2O3} single crystals investigated with {Brillouin}–{Mandelstam} light scattering spectroscopy},
    volume = {127},
    issn = {0003-6951},
    url = {https://doi.org/10.1063/5.0284006},
    doi = {10.1063/5.0284006},
    number = {5},
    urldate = {2026-07-17},
    journal = {Applied Physics Letters},
    author = {Wright, Dylan and Guzman, Erick and Bijoy, Md Sabbir Hossen and Wilson, Richard B. and Mudiyanselage, Dinusha Herath and Fu, Houqiang and Kargar, Fariborz and Balandin, Alexander A.},
    month = aug,
    year = {2025},
    pages = {052201},
}

@article{alkandari_anisotropic_2025,
    title = {Anisotropic anharmonicity dictates the thermal conductivity of β -{Ga} 2 {O} 3},
    volume = {111},
    issn = {2469-9950, 2469-9969},
    url = {https://link.aps.org/doi/10.1103/PhysRevB.111.094308},
    doi = {10.1103/PhysRevB.111.094308},
    language = {en},
    number = {9},
    urldate = {2026-08-27},
    journal = {Physical Review B},
    author = {Alkandari, Abdulaziz and Han, Zherui and Guo, Ziqi and Beechem, Thomas E. and Ruan, Xiulin},
    month = mar,
    year = {2025},
    pages = {094308},
}

@article{zhou_abnormal_2025,
    title = {Abnormal thermal conductivity increase in \textit{β}-{Ga2O3} by an unconventional bonding mechanism using machine-learning potential},
    volume = {52},
    issn = {2542-5293},
    url = {https://www.sciencedirect.com/science/article/pii/S2542529325000331},
    doi = {10.1016/j.mtphys.2025.101677},
    urldate = {2026-08-27},
    journal = {Materials Today Physics},
    author = {Zhou, Wu-Xing and Wu, Cheng-Wei and Cao, Hao-Ran and Zeng, Yu-Jia and Xie, Guofeng and Zhang, Gang},
    month = mar,
    year = {2025},
    pages = {101677},
}

@article{carini_spectral_2026,
    title = {Spectral {Tuning} of {Hyperbolic} {Shear} {Polaritons} in {Monoclinic} {Gallium} {Oxide} via {Isotopic} {Substitution}},
    volume = {38},
    copyright = {© 2026 The Author(s). Advanced Materials published by Wiley-VCH GmbH},
    issn = {1521-4095},
    url = {https://onlinelibrary.wiley.com/doi/abs/10.1002/adma.202514561},
    doi = {10.1002/adma.202514561},
    language = {en},
    number = {11},
    urldate = {2026-08-30},
    journal = {Advanced Materials},
    author = {Carini, Giulia and Pradhan, Mohit and Gelžinytė, Elena and Ardenghi, Andrea and Dixit, Saurabh and Obst, Maximilian and Senarath, Aditha S. and Mueller, Niclas S. and Álvarez-Pérez, Gonzalo and Diaz-Granados, Katja and Kowalski, Ryan A. and Niemann, Richarda and Kaps, Felix G. and Wetzel, Jakob and Iyer, Raghunandan Balasubramanyam and Mazzolini, Piero and Schubert, Mathias and Klopf, J. Michael and Margraf, Johannes T. and Bierwagen, Oliver and Wolf, Martin and Reuter, Karsten and Eng, Lukas M. and Kehr, Susanne C. and Caldwell, Joshua D. and Carbogno, Christian and Folland, Thomas G. and Wagner, Markus R. and Paarmann, Alexander},
    year = {2026},
    pages = {e14561},
}

@article{hajizadeh_selftrapped_2026,
    title = {Self‐{Trapped} {Hole} {Migration} and {Defect}‐{Mediated} {Thermal} {Quenching} of {Luminescence} in α‐ and β‐{Ga}$_{\textrm{2}}$ {O}$_{\textrm{3}}$},
    volume = {36},
    issn = {1616-301X, 1616-3028},
    url = {https://advanced.onlinelibrary.wiley.com/doi/10.1002/adfm.202517876},
    doi = {10.1002/adfm.202517876},
    language = {en},
    number = {13},
    urldate = {2026-08-30},
    journal = {Advanced Functional Materials},
    author = {Hajizadeh, Nima and Frodason, Ymir K. and Petersen, Clemens and Janzen, Benjamin M. and Choi, Luca Sung‐Min and Bernhardt, Nils and Nippert, Felix and Galazka, Zbigniew and Varley, Joel B. and Von Wenckstern, Holger and Vines, Lasse and Wagner, Markus R.},
    month = feb,
    year = {2026},
    pages = {e17876},
}

@misc{tornatzky_complete_2026,
    url = {http://arxiv.org/abs/2607.21045},
    doi = {10.48550/arXiv.2607.21045},
    urldate = {2026-08-30},
    publisher = {arXiv},
    author = {Tornatzky, Hans and Rose, Jonas and Meißner, Moritz and Janzen, Benjamin M. and Galazka, Zbigniew and Reparaz, Juan Sebastián and Ramsteiner, Manfred E. and Wagner, Markus R.},
    month = jul,
    year = {2026},
    note = {arXiv:2607.21045},
}

@article{villora_electrical_2008,
    title = {Electrical conductivity and lattice expansion of β-{Ga2O3} below room temperature},
    volume = {92},
    issn = {0003-6951},
    url = {https://doi.org/10.1063/1.2910770},
    doi = {10.1063/1.2910770},
    number = {20},
    urldate = {2026-08-30},
    journal = {Applied Physics Letters},
    author = {Víllora, Encarnación G. and Shimamura, Kiyoshi and Ujiie, Takekazu and Aoki, Kazuo},
    month = may,
    year = {2008},
    pages = {202118},
}

@article{su_unveiling_2026,
    title = {Unveiling thermal transport properties of defective \${\textbackslash}ensuremath\{{\textbackslash}beta\}{\textbackslash}ensuremath\{-\}{\textbackslash}mathrm\{{G}\}\{{\textbackslash}mathrm\{a\}\}\_\{2\}\{{\textbackslash}mathrm\{{O}\}\}\_\{3\}\$ through machine learning potentials},
    volume = {10},
    url = {https://link.aps.org/doi/10.1103/fdrt-2chf},
    doi = {10.1103/fdrt-2chf},
    number = {4},
    urldate = {2026-08-31},
    journal = {Physical Review Materials},
    publisher = {American Physical Society},
    author = {Su, Yang and Yan, Jin and Yu, Meiyang and Shen, Chen and Fu, Yuhao and Zhou, Tianhang and Zhang, Lijun},
    month = apr,
    year = {2026},
    pages = {044601},
}

@article{wang_dissimilar_2024,
    title = {Dissimilar thermal transport properties in κ-{Ga2O3} and β-{Ga2O3} revealed by homogeneous nonequilibrium molecular dynamics simulations using machine-learned potentials},
    volume = {135},
    issn = {0021-8979},
    url = {https://doi.org/10.1063/5.0185854},
    doi = {10.1063/5.0185854},
    number = {6},
    urldate = {2026-08-31},
    journal = {Journal of Applied Physics},
    author = {Wang, Xiaonan and Yang, Jinfeng and Ying, Penghua and Fan, Zheyong and Zhang, Jin and Sun, Huarui},
    month = feb,
    year = {2024},
    pages = {065104},
}

@article{guo_revisiting_2026,
    title = {Revisiting the thermal conductivity of \${\textbackslash}ensuremath\{{\textbackslash}beta\}{\textbackslash}text\{-{G}\}\{{\textbackslash}mathrm\{a\}\}\_\{2\}\{{\textbackslash}mathrm\{{O}\}\}\_\{3\}\$: {High}-precision benchmarking and the role of interface in time-domain thermoreflectance},
    volume = {10},
    shorttitle = {Revisiting the thermal conductivity of \${\textbackslash}ensuremath\{{\textbackslash}beta\}{\textbackslash}text\{-{G}\}\{{\textbackslash}mathrm\{a\}\}\_\{2\}\{{\textbackslash}mathrm\{{O}\}\}\_\{3\}\$},
    url = {https://link.aps.org/doi/10.1103/myvg-7t2h},
    doi = {10.1103/myvg-7t2h},
    number = {8},
    urldate = {2026-09-02},
    journal = {Physical Review Materials},
    publisher = {American Physical Society},
    author = {Guo, Mengchuan and Liu, Yiyuan and Mu, Wenxiang and Sun, Bo},
    month = aug,
    year = {2026},
    pages = {084603},
}

@incollection{Klemens1958,
  author    = {Klemens, P. G.},
  title     = {Thermal Conductivity and Lattice Vibrational Modes},
  booktitle = {Solid State Physics},
  editor    = {Seitz, Frederick and Turnbull, David},
  volume    = {7},
  pages     = {1--98},
  publisher = {Academic Press},
  address   = {New York},
  year      = {1958},
  doi       = {10.1016/S0081-1947(08)60551-2}
}

@incollection{Slack1979,
  author    = {Slack, Glen A.},
  title     = {The Thermal Conductivity of Nonmetallic Crystals},
  booktitle = {Solid State Physics},
  editor    = {Ehrenreich, Henry and Seitz, Frederick and Turnbull, David},
  volume    = {34},
  pages     = {1--71},
  publisher = {Academic Press},
  address   = {New York},
  year      = {1979},
  doi       = {10.1016/S0081-1947(08)60359-8}
}

@article{maznev_onset_2011,
    title = {Onset of nondiffusive phonon transport in transient thermal grating decay},
    volume = {84},
    issn = {1098-0121},
    url = {https://doi.org/10.1103/PhysRevB.84.195206},
    doi = {10.1103/PhysRevB.84.195206},
    number = {19},
    urldate = {2026-09-24},
    journal = {Physical Review B},
    author = {Maznev, A. A. and Johnson, Jeremy A. and Nelson, Keith A.},
    month = nov,
    year = {2011},
    pages = {195206},
}

@article{johnson_direct_2013,
    title = {Direct Measurement of Room-Temperature Nondiffusive Thermal Transport Over Micron Distances in a Silicon Membrane},
    volume = {110},
    issn = {0031-9007},
    url = {https://doi.org/10.1103/PhysRevLett.110.025901},
    doi = {10.1103/PhysRevLett.110.025901},
    number = {2},
    urldate = {2026-09-24},
    journal = {Physical Review Letters},
    author = {Johnson, Jeremy A. and Maznev, A. A. and Cuffe, John and Eliason, Jeffrey K. and Minnich, Austin J. and Kehoe, Timothy and Sotomayor Torres, Clivia M. and Chen, Gang and Nelson, Keith A.},
    month = jan,
    year = {2013},
    pages = {025901},
}

@article{maznev_generation_2021,
    title = {Generation and detection of 50 GHz surface acoustic waves by extreme ultraviolet pulses},
    volume = {119},
    issn = {0003-6951},
    url = {https://doi.org/10.1063/5.0060575},
    doi = {10.1063/5.0060575},
    number = {4},
    urldate = {2026-09-24},
    journal = {Applied Physics Letters},
    author = {Maznev, A. A. and Mincigrucci, Riccardo and Bencivenga, Filippo and Unikandanunni, V. and Capotondi, Flavio and Chen, Gang and Ding, Zhiwei and Duncan, R. A. and Foglia, Laura and Izzo, M. G. and Masciovecchio, Claudio and Martinelli, Andrea and Monaco, Giulio and Pedersoli, Emanuele and Bonetti, Stefano and Nelson, Keith A.},
    month = jul,
    year = {2021},
    pages = {044102},
}

\clearpage

\setcounter{figure}{0}
\setcounter{subsection}{0}
\renewcommand{\thefigure}{S\arabic{figure}}
\renewcommand{\thesubsection}{S\arabic{subsection}}

\section*{Supporting Information}

\subsection{Coordinate system and crystallographic directions}

\begin{figure}[thb]
\includegraphics[scale=0.6]{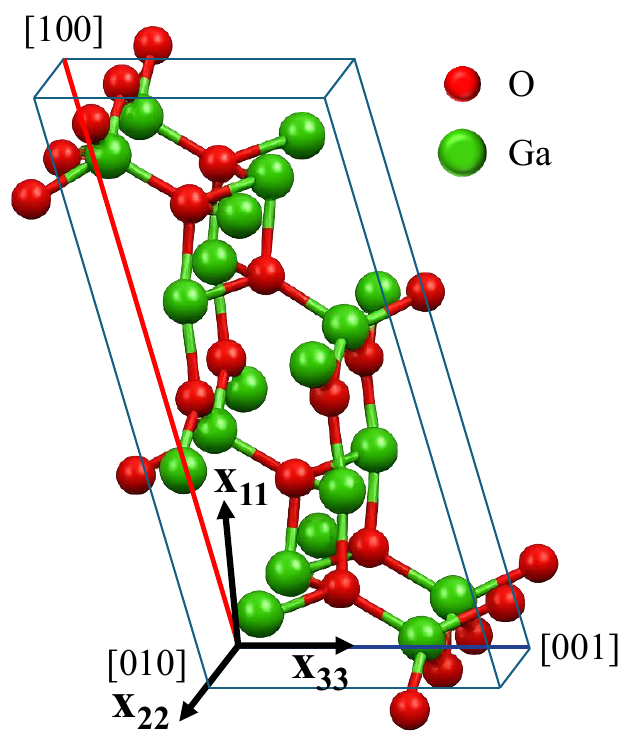}
\caption{Three-dimensional Cartesian coordinate system with monoclinic $\beta$-Ga$_2$O$_3$ unit cell used to describe the thermal-conductivity tensor. The [010] and [001] directions are parallel to $x_{22}$ and $x_{33}$, respectively. The [100] direction lies in the $x_{11}$--$x_{33}$ plane and forms an angle of $13.7^\circ$ with the $x_{11}$ axis.}
\label{fig_crystal_structure}
\end{figure}

We use a symmetry-adapted orthonormal Cartesian basis $(x_{11},x_{22},x_{33})$ in which $x_{22}\parallel[010]$ and $x_{33}\parallel[001]$. The $[100]$ direction lies in the $x_{11}$--$x_{33}$ plane and forms a geometric tilt angle $\alpha=13.7^{\circ}$ with the positive $x_{11}$ axis toward the negative $x_{33}$ direction (see Figure~\ref{fig_crystal_structure}). The corresponding unit vectors are therefore
\begin{align}
\hat{\mathbf{n}}_{[010]} &= (0,1,0),\\
\hat{\mathbf{n}}_{[001]} &= (0,0,1),\\
\hat{\mathbf{n}}_{[100]} &= (\cos\alpha,0,-\sin\alpha).
\end{align}

In this basis, the thermal-conductivity tensor is
\begin{equation}
\boldsymbol{\kappa} =
\begin{pmatrix}
\kappa_{11} & 0 & \kappa_{13}\\
0 & \kappa_{22} & 0\\
\kappa_{13} & 0 & \kappa_{33}
\end{pmatrix}.
\label{eq:kappa_tensor_SI}
\end{equation}

Using $\kappa_{\mathrm{equiv}}(\mathbf{n})
=\mathbf{n}^{\mathrm{T}}\boldsymbol{\kappa}\mathbf{n}$, the conductivities
along the principal crystallographic directions are
\begin{align}
\kappa_{[010]} &= \kappa_{22},\\
\kappa_{[001]} &= \kappa_{33},\\
\kappa_{[100]} &=
\kappa_{11}\cos^2\alpha
-\kappa_{13}\sin(2\alpha)
+\kappa_{33}\sin^2\alpha.
\label{eq:kappa_100_SI}
\end{align}

For the angular analysis of each sample plane, we use a local two-dimensional Cartesian coordinate system. If $\theta$ denotes the angle measured from the local $x$ axis, the in-plane conductivity is
\begin{equation}
\kappa(\theta)
=
\kappa_{xx}\cos^2\theta
+
2\kappa_{xy}\sin\theta\cos\theta
+
\kappa_{yy}\sin^2\theta.
\label{eq:kappa_theta_SI}
\end{equation}
The angle $\theta$ is a local scan angle and is distinct from the geometric tilt
angle $\alpha$.

\clearpage

\subsection{Elastic tensor and Young's moduli}

The elastic anisotropy of \betagao{} is reflected in strong directional dependencies of the Young's moduli ($E_{1}\ll E_{2},E_{3}$), longitudinal-moduli ($C_{11}\ll C_{22}, C_{33}$), as well as shear-moduli ($C_{44} \ll C_{55}, C_{66}$)\cite{adachi2018unusual,Grashchenko2018,Luan2019,Furthmueller2016}. Lower values of $E_{11}$ or the longitudinal modulus $C_{11}$ indicate the strongest elastic anisotropy along the [100] axis, denoting that the material is more compressible in this direction\cite{adachi2018unusual,Luan2019}. Small values of the shear modulus $C_{44}$ would commonly coincide with bond angle change in order for the structure to deform without volume change\cite{adachi2018unusual}. In the case of \betagao{}, however, the shear deformation in the (100)-plane is achieved with smaller angle change by the O-Ga-O bond changing into a parallelogram.\cite{adachi2018unusual}.

Conventionally, the elastic constants may be determined using the plate-thickness resonance\cite{Koga1958} or pulse-echo\cite{Smith1971} methods. However, these methods require the solving of at least 13 equations by measurements of different samples and crystallographic orientations. More recent contact-free approaches include resonant ultrasound spectroscopy coupled with laser-Doppler interferometry (RUS/LDI)\cite{adachi2018unusual,Ogi2002,Ogi2006,Nakamura2012} or polarized Brillouin Light Scattering (BLS)\cite{Graczykowski2015,graczykowski2019brillouin,speziale2014brillouin,vacher1972brillouin,babacic2021thickness} of acoustic phonons in the GHz range as used in this work. 

For small deformations, for which the stress--strain relation remains in the linear regime, the elastic properties of monoclinic \betagao{} are described by 13 independent second-order elastic constants (SOECs). In Voigt notation, these constants are represented by the following $6\times6$ symmetric matrix~\cite{nye1985physical}:
\begin{equation}
C_{IJ} = \begin{pmatrix}
C_{11} & C_{12} & C_{13} & 0 & C_{15} & 0 \\   
C_{12} & C_{22} & C_{23} & 0 & C_{25} & 0 \\   
C_{13} & C_{23} & C_{33} & 0 & C_{35} & 0 \\   
0 & 0 & 0 & C_{44} & 0 & C_{46} \\   
C_{15} & C_{25} & C_{35} & 0 & C_{55} & 0 \\   
0 & 0 & 0 & C_{46} & 0 & C_{66} 
\end{pmatrix}
\label{eqSOE}
\end{equation}

To determine a complete set of SOECs for \betagao{}, we applied Brillouin Light Scattering (BLS)\cite{vacher1972brillouin,speziale2014brillouin,graczykowski2019brillouin,babacic2021thickness} to measure speed of hypersonic acoustic phonons as a function of the direction of propagation.  In the long-wavelength limit the phase velocity $v$ of acoustic phonons can be derived from Christoffel’s equation:
\begin{equation}
|C_{ijmn}l_jl_m-\delta_{in}\rho v^2|=0
\label{eqChris}
\end{equation}
where the direction cosines $l_j$ define the propagation direction of the acoustic wave vector, $\rho$ denotes the mass density, $\delta_{in}$ is the Kronecker delta, and the Einstein summation convention is used. Equation (\ref{eqChris}) has three solutions, which correspond to three acoustic waves/phonons of different polarizations with respect to the acoustic wave vector \textbf{q}. For elastically isotropic materials and high symmetry directions of lower symmetry crystals, those polarizations are purely longitudinal or transverse. For an arbitrary direction of an anisotropic material solutions of Eq. \ref{eqChris} correspond to acoustic waves of mixed polarizations, i.e., (one) quasi-longitudinal and (two) quasi-transverse.

To determine the full elastic tensor $C_{IJ}$  we applied a numerical fitting procedure (reduced $\chi^2$) based on Eq.~\eqref{eqChris}, using 13 fitting parameters adjusted simultaneously for all measured planes and directions. The calculated SOE constants were checked with respect to the crystal stability criteria\cite{mouhat2014necessary} and are listed with the Young's moduli calculated for three main crystallographic directions in Table~\ref{tab:Cij}. The elastic constants obtained from the BLS results follow the relations reported in calculated and measured data from the literature, namely $C_{22}>C_{33}>C_{11}$, $C_{66}>C_{55}>C_{44}$, $C_{13}>C_{12}>C_{23}$, and $C_{15}<0$~\cite{miller2017numerical,doi:10.1063/1.4812323,Grundmann2017,adachi2018unusual}.

\begin{table*}[tb]
\caption{Non-zero independent components of the elastic tensor $C_{KL}$ (GPa) and Young's moduli $E$ (GPa) for three main directions determined from BLS experimental data. }
\label{tab:Cij}
\begin{tabular}{cccccccc}
\toprule[1pt]
$C_{11}$ & $C_{22}$ & $C_{33}$ & $C_{44}$ & $C_{55}$ & $C_{66}$ & $C_{12}$ & $C_{13}$ \\
\midrule[0.5pt]
$220.2 \pm 7.9$ & $343.9 \pm 6.1$ & $328.7 \pm 7.6$ & $47.1 \pm 3$ & $85.3 \pm 2.1$ & $101.7 \pm 5.1$ &   $119.4 \pm 6.9$ &   $141.0 \pm 3.5$ \\
\midrule[0.5pt]
$C_{15}$ & $C_{23}$ & $C_{25}$ & $C_{35}$ & $C_{46}$ & $E_{[100]}$ & $E_{[010]}$ & $E_{[001]}$  \\
\midrule[0.5pt]
$-3.7 \pm 1.3$ &   $76.8 \pm5.3$ &   $-4.0 \pm 2.1$ &  $17.9 \pm 1.2$ &   $4.7 \pm 2.3$ &   $151.6 \pm 7.2$ &  $279.1 \pm 7.7$ &   $233.6 \pm 7.9$\\
\bottomrule[1pt]
\end{tabular}
\end{table*}

\clearpage

\subsection{Phonon velocities and mean free paths from aiGK}
\label{sec:SI_vel_mfp}

\begin{figure*}[h]
\centering
\includegraphics[width=\textwidth]{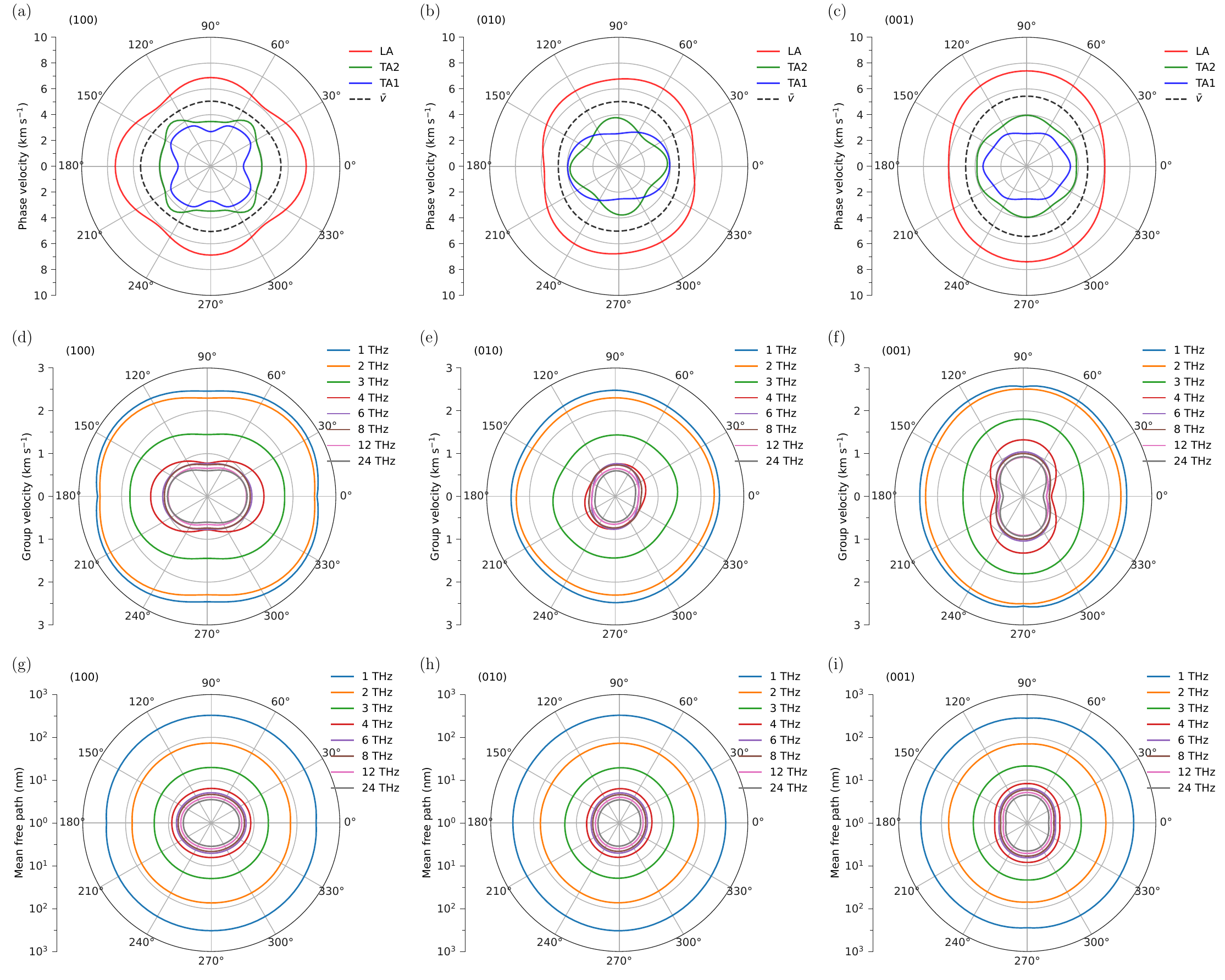}
\caption{(a)-(c) Phase velocities close to $\Gamma$ for the longitudinal~(LA) and transverse~(TA) acoustic modes calculated from the force constants extracted from the \textit{ab initio} molecular dynamics trajectories. The dashed line denotes the cubic mean $\bar{v}(\phi)$ defined in the main text. (d)-(f) Cubic mean group velocities and (g)-(i) MFPs, accumulated over all phonon modes up to the indicated frequency, see Eq.~(\ref{eq:cubicmean}). Phonon mean free paths are extracted from the \textit{ab initio} Green-Kubo calculations.}
\label{sv_gv_mfp}
\end{figure*}

The phase velocities of the acoustic modes shown in Fig.~\ref{sv_gv_mfp}(a)-(c) are obtained by averaging the norm of the group-velocity vectors for wave vectors that lie near the $\Gamma$ point and are oriented along the specified real-space directions. In this limit, group and phase velocity coincide, so that these velocities are directly comparable to the BLS data shown in Fig.~1(d)-(f). The dashed line reflects the cubic mean $\bar{v}(\phi)=\left[\left(v_{L}^3(\phi)+v_{T1}^3(\phi)+v_{T2}^3(\phi)\right)/3\right]^{1/3}$ over the three acoustic branches. In the high-temperature limit, the contribution of an acoustic branch to $\kappa$ scales approximately as $v_j^3$~\cite{Klemens1958,Slack1979}, so the cubic mean provides an estimate of the combined contribution of the three branches along $\phi$.

The same cubic mean is applied to the accumulated group velocity shown in Fig.~\ref{sv_gv_mfp}(d)-(f) and to the accumulated MFP shown in Fig.~\ref{sv_gv_mfp}(g)-(i), so that both are directly comparable to $\bar{v}(\phi)$. These averages are calculated via
\begin{equation}
\bar{X}(\omega_0,\mathbf{n})=
\left(\frac{1}{N_j(\omega_0)}\sum_{j}
\left[\frac{\sum_{\vec{q}}
|\vec{X}(\vec{q},j)\cdot\mathbf{n}|\;
\Theta\bigl(\omega_0-\omega(\vec{q},j)\bigr)}
{\sum_{\vec{q}}
\Theta\bigl(\omega_0-\omega(\vec{q},j)\bigr)}\right]^{3}
\right)^{1/3},
\label{eq:cubicmean}
\end{equation}
with $\vec{X}(\vec{q},j)$ being the group velocity $\vec{v}_g(\vec{q},j)$ or the MFP $\vec{\Lambda}(\vec{q},j)$, $\mathbf{n}$ being the in-plane direction, $\omega_0$ being the cutoff frequency, $\Theta$ being the Heaviside step function, and $N_j(\omega_0)$ being the number of branches that carry modes below $\omega_0$. Since $N_j$ depends on $\omega_0$, a branch enters the cubic mean over the branches only once it carries modes below the cutoff.

\clearpage

\subsection{Angular-resolved in-plane thermal conductivity (ATT)}

\begin{figure*}[h]
\centering
\includegraphics[width=\textwidth]{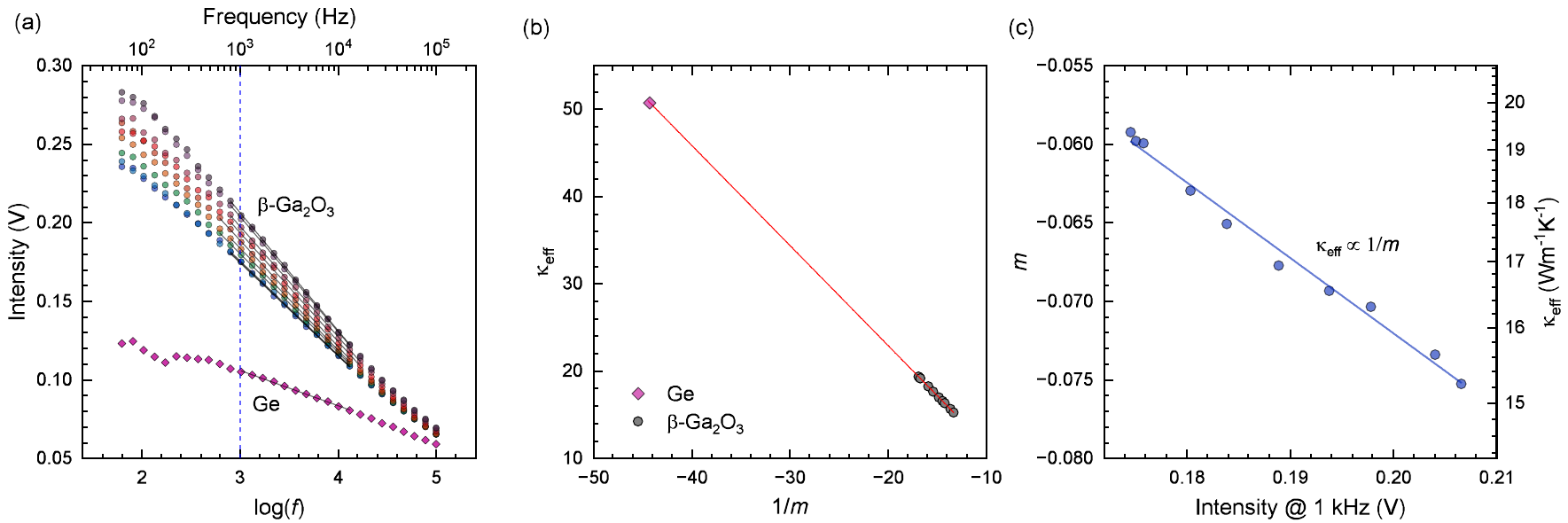}\\
\vspace{0.2cm}
\includegraphics[width=\textwidth]{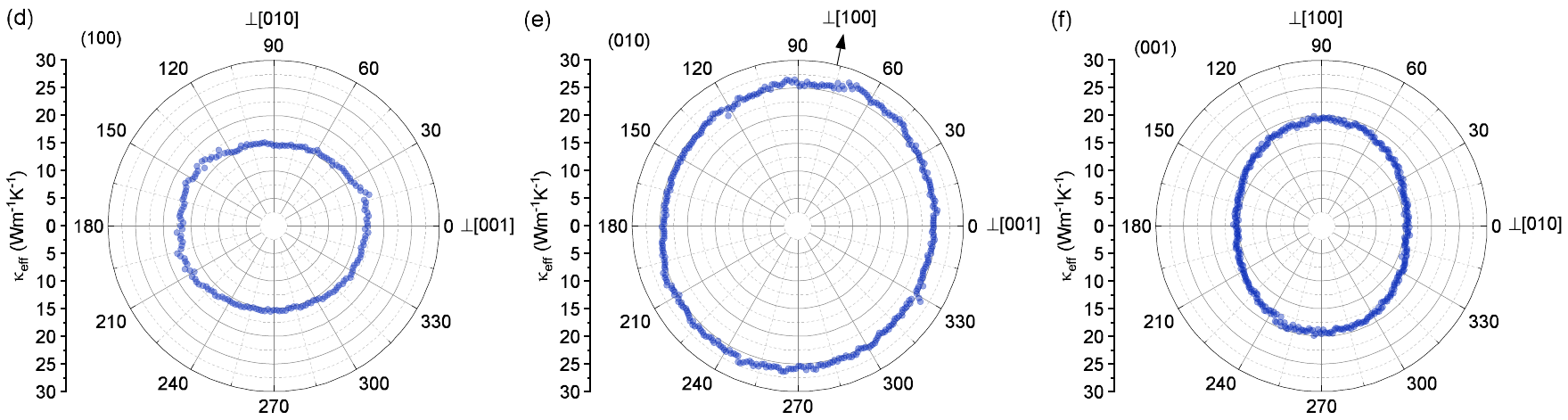}
\caption{(a)-(c) ATT calibration procedure. (a) Variation of the thermoreflectance signal amplitude as a function of the logarithm of the modulation frequency measured by the ATT experiment. The plot includes results for a Ge wafer and (001) $\beta$-Ga$_2$O$_3$ at multiple angles, together with linear fits to each data set using the relation \textit{y=mx+b}, where both the slope \textit{m} and the intercept \textit{b} are treated as free fitting parameters. (b) Linear relationship between the fitted slope m and the effective thermal conductivity $\kappa_{\mathrm{eff}}$. (c) Slope \textit{m} of the fitted $\beta$-Ga$_2$O$_3$ data in (a) vs. signal amplitude at 1 kHz. The right $y$-axis indicates the equivalent thermal conductivity corresponding to the slope. (d–f) Angular-dependent effective thermal conductivity $\kappa_{\mathrm{eff}}$ obtained from ATT scans of samples with (100), (010), and (001) plane orientations, respectively. The heat-flow directions corresponding to the starting angle and other main crystallographic axes are indicated in each polar plot.}
\label{fig_att_si_2}
\end{figure*}

Anisotropic thermoreflectance thermometry (ATT) was employed to characterize the thermal conductivity anisotropy of the (100), (010), and (001) $\beta$-Ga$_2$O$_3$ samples. 
In this method, line-shaped laser spots (approximately 1~mm long and 10~µm wide) were used to heat and probe the sample surface, enabling measurement of the effective thermal conductivity component perpendicular to the line. 
As described in Ref.~\cite{Perez2022}, the effective thermal conductivity is the geometric mean of the cross-plane thermal conductivity ($\kappa_{\perp}$) and the in-plane thermal conductivity component perpendicular to the line ($\kappa_{\mathrm{in}}$):
\begin{equation}
\kappa_{\mathrm{eff}} = \sqrt{\kappa_{\perp}\,\kappa_{\mathrm{in}}}.
\end{equation}

It can be obtained from a linear fit of the thermoreflectance amplitude $A$ versus $\log_{10} f$ (Figure~\ref{fig_att_si_2}b):
\begin{equation}
\kappa_{\mathrm{eff}}
= k
\left( \frac{\partial A}{\partial \log_{10} f} \right)^{-1},
\end{equation}
where $k$ is an instrument constant that accounts for factors such as the absorbed and reflected laser power, the thermoreflectance coefficient of the transducer, and the overall gain of the electronic readout chain, and $\partial A / \partial \log_{10} f$ is obtained from the slope \textit{m} of the linear fit in Figure~\ref{fig_att_si_2}(b). Because $k$ is difficult to determine directly, the system was calibrated using a germanium wafer with the known thermal conductivity of $51~\mathrm{W\,m^{-1}\,K^{-1}}$  under identical conditions. The resulting calibration curve, which relates the thermal conductivity to the fitted slope \textit{m}, is shown in Figure~\ref{fig_att_si_2}(b).

A modulation frequency of 1~kHz was then selected for the in-plane angular scans. At this frequency the measurement remains in the linear regime and the amplitude exhibits high slope sensitivity. 
The fitted slopes from Figure~\ref{fig_att_si_2}(b) and the corresponding amplitudes at 1~kHz are plotted in Figure~\ref{fig_att_si_2}(c), establishing the correlation between the 1~kHz amplitude and the effective thermal conductivity $\kappa_{\mathrm{eff}}$. Using this calibrated relationship, we obtained the $\kappa_{\mathrm{eff}}$ of the (100), (010), and (001) $\beta$-Ga$_2$O$_3$ samples from in-plane angular amplitude scans at 1~kHz and plotted them in Figures~\ref{fig_att_si_2}(d)–(f).

Taking the (100) sample in Figure~\ref{fig_att_si_2}(d) as an example, the effective thermal conductivity is
\begin{equation}
\kappa_{\mathrm{eff},(100)}(\theta)
= \sqrt{\kappa_{\perp,(100)}\,\kappa_{\mathrm{in},(100)}(\theta)},
\end{equation}
where $\kappa_{\perp,(100)}$ is the cross-plane conductivity along the direction normal to the (100) plane (tilted by $13.7^{\circ}$ from [100], see Figure~\ref{fig_crystal_structure}), and $\kappa_{\mathrm{in},(100)}(\theta)$ is the angle-dependent in-plane conductivity. 
Thus, the in-plane thermal conductivity plotted in Figure~1(a) is calculated as
\begin{equation}
\kappa_{\mathrm{in},(100)}(\theta)=\frac{\kappa_{\mathrm{eff},(100)}^{2}(\theta)}{\kappa_{\perp,(100)}}.
\end{equation}

The line-shaped spot was aligned parallel to [001] at the start of the scan. 
At $\theta=0^{\circ}$, the in-plane component of $\kappa_{\mathrm{eff},(100)}$ is perpendicular to [001] in the (100) plane, i.e., parallel to [010]. 
As shown in Figure~\ref{fig_crystal_structure}, [010] corresponds to $x_{22}$; therefore
\begin{equation}
\kappa_{\mathrm{in},(100)}(0^{\circ})=\kappa_{[010]}=\kappa_{22}.
\end{equation}

The local coordinate axes and the zero-angle conventions for the three sample orientations are indicated in Figure~\ref{fig_crystal_structure} and the corresponding ATT scans in Figure~1(a)-(c). Equation~\eqref{eq:kappa_theta_SI} was applied in each local coordinate system to fit the angular dependence and determine the independent tensor components.

\clearpage

\subsection{In-plane thermal diffusivity measurements (BO-FDTR)}

\begin{figure}[thb]
\includegraphics[scale=1]{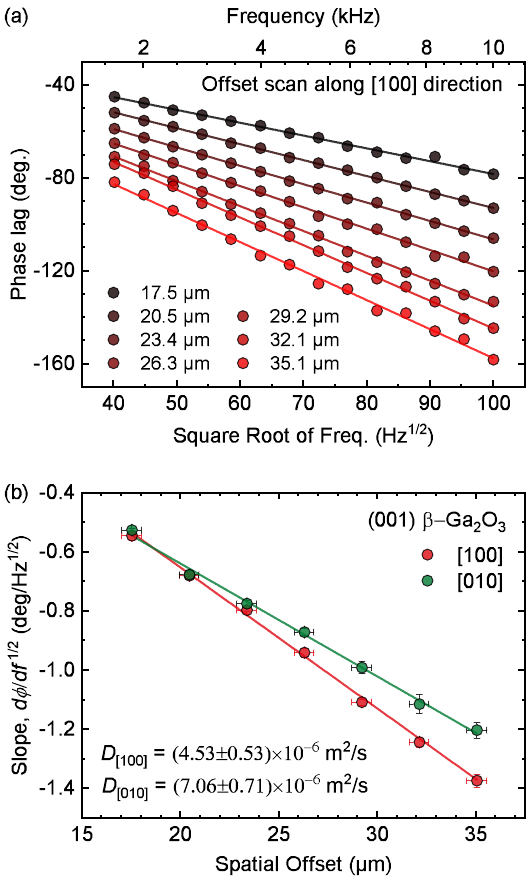}
\caption{(a) Phase lag vs. square root of the modulation frequency for (001)-oriented \mbox{$\beta-$Ga$_{2}$O$_{3}$} with different spatial offsets along the [100] direction. (b) Slope obtained from line fitting $\phi$ vs. $f^{1/2}$ for each spatial offset along [100] and [010] direction.}
\label{fig_offset}
\end{figure}

To further verify the accuracy of the thermal conductivity values obtained by ATT, we employed the recently developed beam-offset frequency-domain thermoreflectance (BO-FDTR) method with a one-dimensional (1D) heat source~\cite{xu_-plane_2023} to measure the in-plane thermal diffusivity of the undoped $\beta$-Ga$_2$O$_3$ (001). By introducing a quasi-1D pump laser spot, this approach enables measurements at large spatial offsets, thereby significantly improving the accuracy in defining the in-plane heat transfer direction.

More importantly, the approximate treatment of the heat transfer model at these large offsets allows the thermal diffusivity ($D$) to be determined directly from measurements of phase lag ($\phi$), modulation frequency ($f$), and spatial offset ($\Delta x$). The relationship between these quantities can be expressed as:
\begin{equation}
\frac{\partial^{2} \phi}{\partial x \,\partial f^{1/2}}
= \sqrt{\frac{\pi}{D}} \,.
\end{equation}
As shown in Figure~\ref {fig_offset}(a), we measured the frequency-dependent phase response of $\beta$-Ga$_2$O$_3$ (001) at different offsets in a specific direction. A linear fit was performed on the data points at each offset in the figure to obtain the slope ($m(\Delta x)$). The range of $\Delta x$ and $f$ in the experiment was chosen to satisfy the linear approximation of the phase response, thus, $m(\Delta x)\approx{\partial \phi}/{\partial f^{1/2}}$. We conducted experiments along the [100] and [010] directions and plotted the $m(\Delta x)$ vs. $\Delta x$ in Figure~\ref {fig_offset}(b). By performing a linear fit to the data in this figure, the values of the slope $m'\approx{\partial^{2} \phi}/{(\partial x \,\partial f^{1/2})}$ for both directions were obtained. The thermal diffusivity was then calculated as {$D=\pi/(m')^2$}. Based on this analysis, the anisotropy ratio of the thermal diffusivities in the two directions shown in the figure is approximately $1.56$, which is consistent with the ratio obtained from the ATT scan. By combining the literature values for density ($\rho = 5961~\mathrm{kg\,m^{-3}}$) and heat capacity ($C_{p} = 492~\mathrm{J\,kg^{-1}\,K^{-1}}$) of $\beta$-Ga$_2$O$_3$~\cite{Ahman1996,King1958} with the measured thermal diffusivity, we calculated the in-plane thermal conductivity to be $13.3~\mathrm{W\,m^{-1}\,K^{-1}}$ along [100] and $20.7~\mathrm{W\,m^{-1}\,K^{-1}}$ along [010], which also shows good agreement with the values obtained from the ATT measurement.

\clearpage

\subsection{Thermal anisotropy in doped crystals}

\begin{figure}[thb]
\includegraphics[scale=0.45]{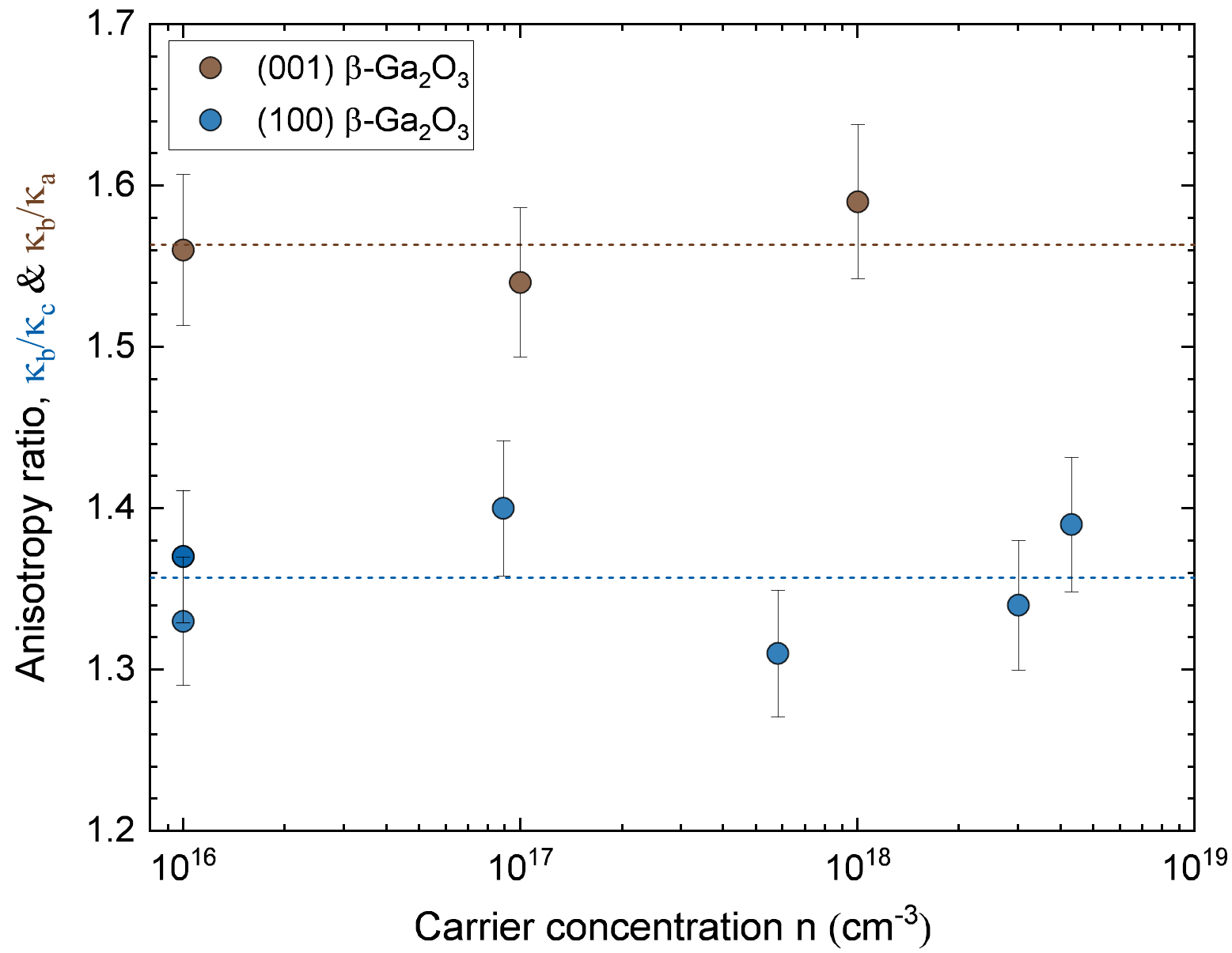}
\caption{Thermal anisotropy ratio of doped $\beta$-Ga$_2$O$_3$ single crystals between the principal axes of the (100) and (001) planes as function of free-carrier concentration.}
\label{fig_k_doping}
\end{figure}

While doping is required to tune electrical properties, it could, in principle, modify the anisotropy of the thermal conductivity by impurity scattering and thermal transport via free carriers. Thus, we investigated the potential influence of free-carrier concentration by Si doping on the in-plane thermal anisotropy. We study a variety of Si doped (100) and (001) \betagao{} samples with free-carrier concentrations between $n=1\cdot10^{16} - 5\cdot10^{18}$ cm$^{-3}$. Figure~\ref{fig_k_doping} illustrates the maximum anisotropy ratios as a function of carrier concentration for both planes.

The results show that the anisotropy ratios remain insensitive, within experimental uncertainty, to variations in impurity and free-carrier concentrations. This finding indicates that, over the investigated doping range, impurity and free-carrier scattering do not measurably affect the phonon transport responsible for the observed thermal anisotropy. The absence of a measurable change is consistent with the interpretation that the dominant anisotropy originates from relatively short-MFP phonons, as independently inferred from the EUV-TTG and aiGK analyses. 

\clearpage

\subsection{Phonon dispersion}

\begin{figure}[thb]
\includegraphics{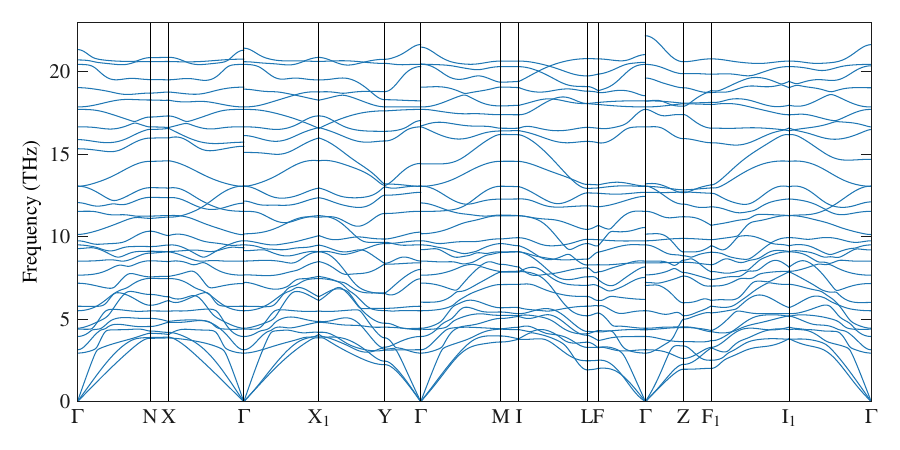}
\caption{Phonon dispersion relation obtained from harmonic force constants fitted to the aiMD trajectory using TDEP.}
\label{fig_phonon_bandstructure}
\end{figure}

The phonon dispersion relation shown in Fig.~\ref{fig_phonon_bandstructure} is obtained from the effective harmonic force constants that are also used for the mode decomposition of the \textit{ab initio} Green-Kubo thermal conductivity $\kappa^{\alpha\beta}$ and for the cubic means in Eq.~(\ref{eq:cubicmean}), see the subsection ``Ab initio Green-Kubo Calculations in the main manuscript for more details. The frequencies are effective harmonic frequencies at 300~K, extending up to 22~THz. The acoustic branches reach the zone boundary between 2 and 4~THz, depending on the direction. The lowest optical branches start at 3~THz at $\Gamma$.

\clearpage

\subsection{Literature overview of thermal conductivities}

\begin{table*}[h]
\centering
\caption{Reported thermal conductivities of $\beta$-Ga$_2$O$_3$ in W\,m$^{-1}$\,K$^{-1}$.}
\label{tab:summary}
\begin{threeparttable}
\renewcommand{\baselinestretch}{1.1}\selectfont
\begin{tabular}{lllrrrrrrr}
\toprule
\textbf{Author}        & \textbf{Year} & \textbf{Method} & \textbf{a-axis} & \textbf{b-axis}  & \textbf{c-axis} & \textbf{b/a} & \textbf{b/c} & \textbf{c/a} & \textbf{Ref.}\\ \midrule
This work        & 2026   & ATT                & 13.6 & 21.2 & 16.0 & 1.6 & 1.3 & 1.2 & \\
Guo              & 2026   & TDTR               & 13.0 & 22.5 & 17.7 & 1.7 & 1.3 & 1.4 & \cite{guo_revisiting_2026}\\ 
Klimm            & 2023   & Laser-flash        & 12.7 & 24.3 & 14.1 & 1.9 & 1.7 & 1.1 & \cite{Klimm2022}\\
Galazka          & 2022   & Laser-flash        & 12.2 & 24.6 & 17.6 & 2.0 & 1.4 & 1.4 & 
\cite{Galazka2022}\\
Jiang            & 2018   & TDTR               &  9.5 & 22.5 & 13.3 & 2.4 & 1.7 & 1.4 & \cite{Jiang2018}\\
Mu               & 2017   & Laser-flash        & 14.9 & 27.9 & 17.9 & 1.9 & 1.6 & 1.2 & \cite{Mu2017}\\
Handwerg         & 2016   & 2$\omega$          & 11.0 & 29.0 & 21.0 & 2.6 & 1.4 & 1.9 & \cite{Handwerg2016}\\
Guo              & 2015   & TDTR               & 10.9 & 27.0 & 14.7 & 2.5 & 1.8 & 1.3 & \cite{Guo2015}\\
\hline
This work        & 2026   & DFT-TDEP + aiGK    &  9.5 & 16.5 & 12.1 & 1.7 & 1.4 & 1.3 & \\
This work        & 2026   & DFT + BTE          & 14.3 & 22.6 & 14.6 & 1.6 & 1.5 & 1.0 & \\
Su               & 2026   & MLP + EMD       & 9.3  & 19.8 & 12.5 & 2.1 & 1.6 & 1.3 &
\cite{su_unveiling_2026}\\
Su               & 2026   & MLP + BTE      & 12.9 & 19.3 & 14.2 & 1.5 & 1.4 & 1.1 & \cite{su_unveiling_2026}\\
Alkandari        & 2025   & DFT-TDEP + BTE     & 15.0 & 27.0 & 19.5 & 1.8 & 1.4 & 1.3 & \cite{alkandari_anisotropic_2025}\\
Zhou             & 2025   & DFT-NEP + MD       & 14.6 & 19.2 & 17.5 & 1.3 & 1.1 & 1.2 & \cite{zhou_abnormal_2025}\\
Tu               & 2025   & DFT + BTE(3+4ph)  & 12.2 & 17.9 & 14.9 & 1.5 & 1.2 & 1.2 & \cite{Tu2025}\\
Safieddine       & 2024   & DFT + BTE         & 14.3 & 29.6 & 20.2 & 2.1 & 1.5 & 1.4 & \cite{Safieddine2024}\\
Rybin            & 2024   & DFT-MTP + BTE     &  9.9 & 17.5 & 12.9 & 1.8 & 1.4 & 1.3 & \cite{Rybin2024}\\
Wang             & 2024   & DFT-NEP + HNEMD   & 10.3 & 19.9 & 12.6 & 1.9 & 1.6 & 1.2 &
\cite{wang_dissimilar_2024}\\
Yang             & 2024   & DFT + BTE         & 10.6 & 18.9 & 13.9 & 1.8 & 1.4 & 1.3 & \cite{Yang2024}\\ 
Chen             & 2023   & DFT-TDEP + BTE    & 10.0 & 23.7 & 12.0 & 2.4 & 2.0 & 1.2 & \cite{Chen2023}\\
Sun              & 2023   & DFT-NEP + MD       &  9.4 & 21.5 & 11.9 & 2.3 & 1.8 & 1.3 & \cite{Sun2023}\\
Li               & 2020   & DFT-NN + EMD        & 10.7 & 20.8 & 12.6 & 1.9 & 1.6 & 1.2 & \cite{Li2020}\\
Yan              & 2018   & DFT + BTE         & 12.7 & 20.0 & 17.8 & 1.6 & 1.1 & 1.4 & \cite{Yan2018}\\
Santia           & 2015   & DFT + BTE         & 16.1 & 21.5 & 21.2 & 1.3 & 1.0 & 1.3 & \cite{Santia2015}\\
\bottomrule
\vspace{-1em}
\end{tabular}
\end{threeparttable}
\begin{minipage}[t]{\textwidth}
\footnotesize
\justifying
\renewcommand{\baselinestretch}{1.0}\selectfont
\noindent
Directions: a-axis, b-axis, c-axis denote thermal conductivity along crystallographic directions [100], [010], [001]; anisotropy ratios defined as $b/a=\kappa_{[010]}/\kappa_{[100]}$, $b/c=\kappa_{[010]}/\kappa_{[001]}$, $c/a=\kappa_{[001]}/\kappa_{[100]}$.\\
Experimental methods: ATT: anisotropic thermoreflectance thermometry;
TDTR: time-domain thermoreflectance; 
$2\omega$, $3\omega$: electrical techniques; 
Laser-flash: laser-flash diffusivity measurements.\\
Theoretical methods: DFT: DFT-derived IFCs;
DFT-TDEP: temperature-dependent effective IFCs from DFT-based \textit{ab initio} MD; 
DFT-NN: DFT-trained neural-network potential; 
DFT-NEP: DFT-trained neuroevolution potential; 
DFT-MTP: DFT-trained moment-tensor potential; 
MLP: machine-learning potential;
MD: molecular dynamics;
EMD: equilibrium molecular dynamics;
HNEMD: homogeneous nonequilibrium molecular dynamics;
aiGK: fully anharmonic \textit{ab initio} Green--Kubo; 
BTE: Boltzmann transport equation, three-phonon scattering; 
BTE(3+4ph): BTE including three- and four-phonon scattering. 
\end{minipage}
\end{table*}

\end{document}